\documentclass[sigconf]{acmart}

\AtBeginDocument{%
  \providecommand\BibTeX{{%
    \normalfont B\kern-0.5em{\scshape i\kern-0.25em b}\kern-0.8em\TeX}}}

\copyrightyear{2024}
\acmYear{2024}
\setcopyright{acmlicensed}
\acmConference[WWW '24] {Proceedings of the ACM Web Conference 2024}{May 13--17, 2024}{Singapore, Singapore.}
\acmBooktitle{Proceedings of the ACM Web Conference 2024 (WWW '24), May 13--17, 2024, Singapore, Singapore}
\acmISBN{979-8-4007-0171-9/24/05}
\acmDOI{10.1145/3589334.3648154}

\usepackage{tabularx}
\usepackage{caption}
\usepackage{subcaption}
\usepackage{booktabs}
\usepackage{soul}
\usepackage{multirow}
\usepackage{makecell}
\usepackage{balance}

\author{Jinyi Ye}
\orcid{0009-0004-7757-1642}
\affiliation{%
  \institution{University of Southern California}
  \department{Information Sciences Institute}
  \streetaddress{4676 Admiralty Way (1001)}
  \city{Marina Del Rey}
  \state{CA}
  \postcode{90292}
  \country{USA}
}
\email{jinyiy@usc.edu}

\author{Luca Luceri}
\orcid{0000-0001-5267-7484}
\affiliation{%
  \institution{University of Southern California}
  \department{Information Sciences Institute}
  \streetaddress{4676 Admiralty Way (1001)}
  \city{Marina Del Rey}
  \state{CA}
  \postcode{90292}
  \country{USA}
}
\email{lluceri@isi.edu}

\author{Julie Jiang}
\orcid{0000-0003-4260-282X}
\affiliation{%
  \institution{University of Southern California}
  \department{Information Sciences Institute}
  \streetaddress{4676 Admiralty Way (1001)}
  \city{Marina Del Rey}
  \state{CA}
  \postcode{90292}
  \country{USA}
}
\email{juliej@isi.edu}

\author{Emilio Ferrara}
\orcid{0000-0002-1942-2831}
\affiliation{%
  \institution{University of Southern California}
  \department{Thomas Lord Department of Computer Science}
  \city{Los Angeles}
  \state{CA}
  \postcode{90089}
  \country{USA}
}
\email{emiliofe@usc.edu}

\begin{document}

\title[Susceptibility to Unreliable Information Sources: Swift Adoption with Minimal Exposure]{Susceptibility to Unreliable Information Sources:\\ Swift Adoption with Minimal Exposure}


\renewcommand{\shortauthors}{}

\begin{abstract}
Misinformation proliferation on social media platforms is a pervasive threat to the integrity of online public discourse. 
Genuine users, susceptible to others' influence, often unknowingly engage with, endorse, and re-share questionable pieces of information, collectively amplifying the spread of misinformation.
In this study, we introduce an empirical framework to investigate users' susceptibility to influence when exposed to unreliable and reliable information sources. Leveraging two datasets on political and public health discussions on Twitter, we analyze the impact of exposure on the adoption of information sources, examining how the reliability of the source modulates this relationship. Our findings provide evidence that increased exposure augments the likelihood of adoption. Users tend to adopt low-credibility sources with fewer exposures than high-credibility sources, a trend that persists even among non-partisan users. 
Furthermore, the number of exposures needed for adoption varies based on the source credibility, with extreme ends of the spectrum (very high or low credibility) requiring fewer exposures for adoption. Additionally, we reveal that the adoption of information sources often mirrors users' prior exposure to sources with comparable credibility levels.
Our research offers critical insights for mitigating the endorsement of misinformation by vulnerable users, offering a framework to study the dynamics of content exposure and adoption on social media platforms.
\end{abstract}





\maketitle

\section{Introduction}






Misinformation and non-credible information are critical challenges in contemporary society, drawing significant scholarly attention in recent years \cite{luceri2024unmasking, singh2022misinformation, linden2022misinformation, chen2022charting}. In the digital age, the rapid dissemination of content on social media platforms has amplified the reach of false narratives, potentially leading to widespread misconceptions and harmful behaviors \cite{walter2020meta}. For example, the 2016 U.S. election witnessed a surge in fake news \cite{guess2018selective}, and misinformation regarding COVID-19 was disseminated at a rate 32.4\% higher than factual content \cite{pennycook2020fighting}. While social bots have previously been shown to have played an important role in sharing questionable content \cite{ferrara2018measuring, ferrara2022twitter}, a substantial amount of recent dissemination of unreliable COVID-19 information were, in fact, from plausibly genuine users \cite{chang2022comparative, teng2022characterizing,nogara2022disinformation}. 

There is a large body of research on the adoption and spread of misinformation and fake news. Most research focuses on answering \textit{why} some people trust unreliable content \cite{ecker2022psychological, pennycook2021psychology, linden2022misinformation}, identifying \textit{who} is most vulnerable to misinformation \cite{aral2012identifying, pennycook2019lazy, calvillo2021individual}, or explaining \textit{how} fake news spread in social networks \cite{vosoughi2018spread, zhou2019network}. Yet much less is understood about the differences between the dynamics of adopting truthful information versus falsehoods. In this work, we bridge this gap in the literature by investigating the distinctions in the information adoption process, considering social network exposure, between credible and non-credible content.



In this paper, we present an empirical study to examine users' susceptibility to influence when exposed to a range of information sources, taking into account three key dimensions: the \underline{credibility} of the information sources they are exposed to, the frequency of \underline{exposure} to the information source, and the likelihood of subsequent \underline{adoption} of the information source. In this work, we look at the information shared on Twitter through URLs to popular media and analyze how the exposure-adoption mechanisms differ for media of varying credibility. We approximate exposure to media content by considering the URLs shared by a user's social connections, and we operationalize adoption as the subsequent sharing of URLs from a particular media by the user.
\subsection*{Contributions of this work}
Guided by our motivation  to investigate the interplay between users' propensity to adopt an information source, the reliability of the source, and the extent of exposure (see Fig. \ref{fig:RQs}),
we formulate and investigate the three following research questions (RQs):

\begin{figure}
    \centering
    \includegraphics[width=6cm]{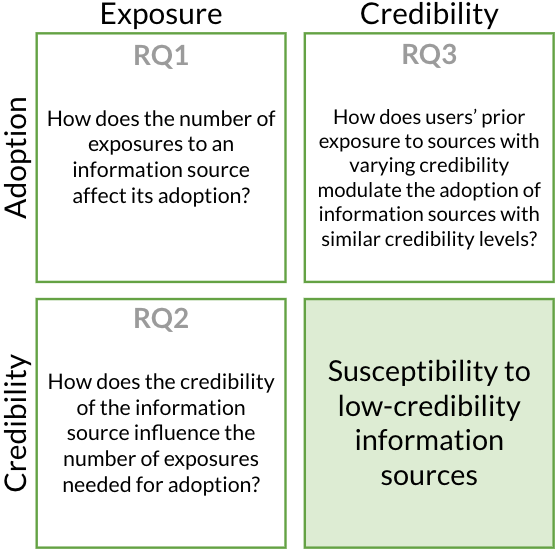}
    \caption{RQs and components of the proposed methodological framework.}
    \label{fig:RQs}
\end{figure}


\begin{itemize}
\item[\textbf{RQ1}:] How does the number of exposures to an information source affect its adoption, and how is this effect modulated by the credibility of the information source?

\item[\textbf{RQ2}:] How does the credibility of the information source influence the number of exposures needed for adoption?

\item[\textbf{RQ3}:] 
How does users’ prior exposure to sources with varying credibility modulate the adoption of information sources with similar credibility levels?
\end{itemize}

Our analysis reveals compelling evidence that increased exposure frequency corresponds to a higher likelihood of source adoption. 
Notably, given the same level of exposure, users are more likely to adopt low-credibility sources, a trend that is robust to potential political bias (RQ1). 
Considering all sources that are adopted, both extremely low- and high-credibility sources require fewer exposures prior to adoption than sources of moderate credibility (RQ2). 
Furthermore, we observe a strong positive correlation between the credibility of sources users are exposed to and those they subsequently adopt (RQ3). Specifically, users are more likely to adopt low-credibility sources if they are exposed to more low-credibility sources. The same trend holds true for users who adopt high-credibility sources; however, they usually require a much greater amount of high-credibility exposure before adoption.  This suggests that the credibility of exposures may have an effect on corresponding adoptions. Our findings remain consistent across two large-scale Twitter datasets covering political and public health discussions. This research offers critical insights into understanding user behaviors and vulnerabilities on social media platforms, especially their tendencies to align with information sources of specific credibility levels, with broader implications for combating misinformation online.

\section{Related Work}
The extensive literature on misinformation and fake news offers several theories about why people fall prey to them \cite{ecker2022psychological,pennycook2021psychology,linden2022misinformation}. People are likely to accept claims from sources that they perceive to be credible \cite{marsh2018believing,traberg2022birds}. If a piece of information is repeated enough times, the increased exposure will make the claim easier to process and thus more likely taken to be true in a process known as the ``illusory truth'' effect \cite{pennycook2018prior,dechene2010truth,wang2016known}. Partisanship or ideological views may also play a role here due to our tendency to seek out and trust information that conforms to our worldviews (confirmation bias) \cite{guess2018selective,moravec2018fake} or believe in messages from sources that are perceived to be similar to ourselves \cite{traberg2022birds}, especially with common political viewpoints \cite{marks2019epistemic,bauer2021believing}. Misinformation that elicits stronger emotional responses could also render it more believable than truth \cite{chou2020considering,baum2021emotional,solovev2022moral}. The novelty of fake news could be another factor leading to their popularity \cite{vosoughi2018spread}. Another theory is that people often forgo careful reasoning for intuitive, lazy reasoning, rendering them gullible to fake news \cite{pennycook2019lazy,pennycook2021psychology}. Even when people have been corrected with accurate information, many will persist in being affected by the false information owing to the continued influence effect \cite{johnson1994sources,walter2020meta}. 

Many studies seek to answer the question: \textit{who} is susceptible to influence? In a large randomized experiment on Facebook, \citet{aral2012identifying} found individual differences in users' susceptibility to influence; for example, younger users are more susceptible than older users, and men are susceptible to influence from women. Politically right-leaning individuals are shown to be more susceptible to misinformation and deception \cite{calvillo2021individual,pennycook2019lazy,chen2021covid,luceri2019red}. In particular, those with a lower belief in science or higher conspiratorial inclinations are more likely to share health misinformation \cite{saling2021no}. These beliefs often manifest through distinct behavioral cues \cite{luceri2018social}, including signals of group identity, compliance with community norms, and distrust towards opposing views  \cite{wang2023identifying}.

Another line of work examines \textit{how} fake news spreads online. Online diffusion patterns of fake news have been found to be distinct from those of truthful information: fake news tends to travel faster, deeper, and broader \cite{vosoughi2018spread}. Further, they have more spreaders and spreaders who engage more actively with fake news, resulting in a larger and denser fake news network \cite{zhou2019network}. These insights have allowed research to discern misinformation from network diffusion patterns alone \cite{zhou2019network,naumzik2022detecting,rosenfeld2020kernel}.

Nevertheless, several critical questions persist: Are users equally susceptible to influence across information sources with different levels of credibility? Does a person's susceptibility to influence change based on the extent of exposure, thus affecting the adoption of low-credibility sources? Is the adoption rate of misinformation faster than that of truthful content? In this study, we aim to address these gaps in the existing literature by introducing a methodological framework that explores the interrelationships between exposure, adoption, and the credibility of a large suite of information sources.


\section{Problem Definition}


\label{sec:measure_user_sus}
Our conceptualization of a user's susceptibility to influence draws from both psychology \cite{laursen2022does, moussaid2013social} and computer science studies \cite{aral2012identifying, hoang2016tracking, zhou2019network, romero2011differences}. Social psychologists posit that individuals adjust their opinions, beliefs, or behaviors due to the influence of others \cite{moussaid2013social}. In this context, susceptibility indicates an openness to peer influence, often leading to conformity behaviors \cite{laursen2022does}. Similarly, computer scientists studying social media define users’ susceptibility as their likelihood to be infected with items propagated by other users \cite{aral2012identifying, hoang2016tracking,saito2008prediction}. Researchers have operationalized susceptibility to misinformation, or ``vulnerability to misinformation'' according to \citet{nikolov2021right}, as the proportion of low-credibility sources shared by a user \cite{zhou2019network} or the probability of adoption given one or more exposures to misinformation \cite{romero2011differences}. Notably, susceptibility is modeled as a \textit{probability}, capturing the fraction of users’ activity driven by social influence \cite{zhou2019network,romero2011differences,goyal2010learning,luceri2018deep}. These overlapping definitions emphasize two main components of susceptibility: \textit{exposure} to a piece of content and the potential subsequent content \textit{adoption}. 

Given these perspectives, in this work, we operationalize users' susceptibility to influence as their likelihood to adopt content they are exposed to. In this study, we consider media and news outlets as pieces of content, examining exposure to and adoption of information sources. Importantly, our notion of susceptibility is measured independently from the credibility of the content, meaning that susceptibility to \textit{misinformation} sources (low-credibility) is measured identically to susceptibility to \textit{factual} (high-credibility) sources. 

\paragraph{Exposure}
Exposure to an information source on social media generally refers to the event wherein a user encounters or views a particular piece of content from that source, regardless of their subsequent interaction (e.g., like, share, comment). Measuring exposure can be challenging because simply viewing content does not result in an observable action. Nonetheless, exposure data could be gathered through self-reported measures like list-based recall surveys \cite{guess2019accurate}, and observational measures such as online browser web tracking \cite{guess2020exposure,moore2023exposure}, and screen activity \cite{reeves2021screenomics}. 

Following previous research, we approximate exposure leveraging observable sharing activities on social media. Existing methods typically consider that users are exposed to the content shared by their followers. 
However, with the latest feed ranking algorithms, users are not only exposed to posts generated by their followed accounts but also by tweets algorithmically curated by the platform’s recommendation systems, resulting in heightened levels of selective exposure \cite{guess2020exposure}. For example, on Twitter, users can be exposed to and retweet content from any account, with or without a follow relationship. To address this challenge, previous work \cite{rao2022partisan, ferrara2015measuring, rao2022retweets, sasahara2021social} relies on users' observed interactions to approximate their exposure to content generated by those they interact with. Specifically, a target user $u_{t}$ is considered exposed to the set of users retweeted by $u_{t}$. 

We further extend this idea by also considering quote and reply interactions as a proxy for exposure. The underlying assumption is that when users engage in actions like re-sharing, replying, or quoting others, they are necessarily exposed to the content generated by those users. Consequently, they are likely to follow these users or come across their content again in the future.
Therefore, we operationalize exposure as follows: If a target user $u_{t}$ interacts with a user $u_{s}$ through a retweet, quote, or reply, then for each set of tweets $\mathcal{X}$ that $u_{s}$ shares after the time of their first observed interaction, we assume that $u_{t}$ is exposed to tweets $\mathcal{X}$. 


\paragraph{Adoption}
In this study, we consider adoption as the act of sharing a URL containing a particular information source, regardless of its credibility and type of tweet, i.e., original tweet, retweet, quoted tweet, or reply tweet.

\section{Data}
For our study, we carefully select two distinct datasets during a specified time frame to provide a robust look at user susceptibility to online content. The first dataset (hereinafter referred to as the \textsc{Election} dataset), gathered by \citet{chen2021election2020}, tracks over one billion tweets related to the 2020 US election beginning from January 1st, 2020. Tweets were collected by tracking messages mentioning the Twitter accounts of Republican and Democratic candidates in the presidential election. This was achieved by using Twitter’s streaming API service (v.1.1), which captured a roughly 1\% sample of all streaming tweets. We limit our study period to tweets that appeared between January 1st, 2020, and June 30th, 2020 
to study roughly six continuous months of users’ online interaction. This subset of data contains 364 million tweets from 14.5 million users.

The second dataset (hereinafter referred to as the \textsc{Covid} dataset), collected by \citet{chen2020tracking}, contains COVID-19 pandemic-related tweets starting from January 21st, 2020. This dataset tracks tweets mentioning specific COVID-related keywords and accounts, offering a comprehensive view of online conversations and public sentiments regarding the pandemic. Similar to the \textsc{Election} dataset, the \textsc{Covid} dataset was directly collected via Twitter’s API, ensuring completeness of the data with no missing tweets due to re-hydration issues. We similarly limit our study period from January 21st, 2020, to June 30th, 2020, which contains 260 million tweets from 35 million users.

Our choice of the \textsc{Election} and \textsc{Covid} datasets for the overlapping six-month time span of the first half of 2020 is underpinned by the significance of two major events surrounding US national politics and the global public health crisis. Analyzing these pivotal events in tandem offers a unique opportunity to explore people's receptivity to information across distinct topical interests. Furthermore, analyzing both datasets ensures the reliability and consistency of our results under different contexts.

\section{Methodology}
\subsection{Categorizing Information Sources}
Tweets often contain URLs from popular media outlets. We can, therefore, use the credibility of the media outlet as a proxy for the credibility of the information source cited by the tweet. Following prior research \cite{rao2022partisan,pierri2023propaganda,luceri2019red}, we assess content credibility by examining pay-level domains within the URLs embedded in tweets. We curate two distinct lists of domains representing low- and high-credibility information sources shared on social media. 

We use the Media Bias/Fact Check (MBFC) website\footnote{\url{https://mediabiasfactcheck.com}} as well as the Iffy Index of Unreliable Sources\footnote{\url{https://iffy.news/index/}} to assess news media credibility. MBFC is an independent news media watchdog that rates news media on a 6-point factuality scale ranging from \textit{Very Low} to \textit{Very High}. The Iffy Index further expands the reliability rating provided by MBFC by categorizing a source as either Conspiracy/Pseudoscience or Questionable Source/Fake News. 
We categorize a total of 1,481 domains from the Iffy Index as sources with low credibility. Within this low-credibility category, we further classify them based on their MBFC factual scores into three distinct subcategories: \textit{Very Low}, \textit{Low}, and \textit{Mixed}. To identify high-credibility sources, we assemble a list of 1,942 media websites that received a \textit{High} rating in terms of credibility by MBFC. These high-credibility sources are also divided into three separate subcategories based on factual scores: \textit{Very High}, \textit{High}, and \textit{Mostly Factual}. Examples of sources representing these credibility categories are detailed in the \textit{Appendix}. 

For every post across the two datasets, we systematically extract, expand, and parse all embedded URLs. We then examine whether the URL is from either a low- or high-credibility domain from the two lists. The \textsc{Election} dataset contains 4.6 million URLs shared amongst 776,000 users,
while the \textsc{Covid} dataset contains 8.8 million URLs distributed by 2.5 million users.
\begin{table}
  \centering
  \small
  \caption{Number of total users, target users, and breakdown across the three user categories in both datasets. \vspace{-0.2cm}}
  \label{user_summary}
  \begin{tabular}{rrr}
  \toprule
    \textbf{\# of Users} & \textbf{\textsc{Election}} & \textbf{\textsc{Covid}}\\
    \midrule
    All users & 775,937 & 2,504,689\\
    Target users & 21,200 & 34,053\\
    \midrule
    ONLY low-credibility & 1,592 & 1,402\\
    ONLY high-credibility & 8,201 & 18,872\\
    Both low- and high-credibility & 11,587 & 13,779\\
    \bottomrule
\end{tabular}
\end{table}
\subsection{Identifying Target Users}\label{sec:identify_target_user}
We conduct additional filtering to identify a subset of \textit{target users}. 
A user must satisfy two criteria to be considered a \textit{target user}. First, we restrict our attention to users who \textit{have not} shared any specified low- or high-credibility sources during the first few months of our study period but did so later on. We refer to this timeframe as a \textit{buffering} period. Driven by our exposure-adoption framework to assess user susceptibility (\textit{cf.} \S\ref{sec:measure_user_sus}),
this approach narrows our pool of users down to 
estimate the extent of exposure target users experienced before they first adopted an information source. 
In line with \citet{bakshy2012role}, we choose a two-month buffer, 
equivalent to one-third of our study period, to balance a consistent \textit{buffering} timeframe with a representative set of eligible target users.

The second criterion poses a minimum threshold on the number of URLs shared to qualify as a \textit{target user}. Following \cite{nikolov2021right}, we only consider users who shared at least ten, not necessarily unique, links from low- and high-credibility sources. This criterion is particularly relevant as our study aims at understanding the differences across different susceptible populations. Users who share fewer than ten links have insufficient data points to assess their susceptibility to influence confidently and are thus excluded from our analysis.

This filtering results in 21,200 and 34,053 users from the \textsc{Election} and \textsc{Covid} datasets, respectively. We categorize these target users into three mutually exclusive groups: those who \textit{only} shared low-credibility sources, those who \textit{only} shared high-credibility sources, and those who shared both. Table \ref{user_summary} provides an overview of the number of target users meeting the specified criteria and their distribution across the three categories. In both datasets, about half of the users share both low- and high-credibility sources. Users who share {only} low-credibility sources represent the smallest group in both datasets, accounting for less than 10\% of the target users.

\subsection{Quantifying User Credibility and Political Orientation}
To further characterize users in these three groups, we introduce an \textit{adoption credibility} metric. This metric gauges the overall credibility of content adopted by a target user. We devise this metric based on \textit{(i)} the credibility of each information source a user adopts, and \textit{(ii)} the frequency of adopting these sources. To quantify the credibility of each information source, we assign each of them a numerical value on a scale from 0 to 1: \textit{Very Low} (0), \textit{Low} (0.2), \textit{Mixed} (0.4), \textit{Mostly Factual} (0.6), \textit{High} (0.8), and \textit{Very High} (1). We calculate 
 user credibility 
 as the weighted average of the credibility scores of the sources that a user shares, with weights corresponding to the sharing frequency.
It is worth noting that the above metric evaluates the credibility of the sources a user adopts. Similarly, we consider the credibility of the sources a user is exposed to and introduce an \textit{exposure credibility} metric, computed as the weighted average of the credibility scores of the sources a user is exposed to. 



Further, as part of our effort to control for biases arising from political partisanship, we also infer the political leaning of every target user in a similar manner. Utilizing the media bias ratings provided by MBFC, we classify information sources on a 7-point political leaning scale: \textit{Extreme Left} (1), \textit{Left} (2), \textit{Left-Center} (3), \textit{Least Biased} (4), \textit{Right-Center} (5), \textit{Right} (6), and \textit{Extreme Right} (7). These scores are then normalized to a 0 to 1 scale. We exclude domains that do not fall into these categories or are not rated by MBFC from our analysis. The political leaning score of every target user is computed as the weighted mean of the political bias scores of the domains they share, weighted by their sharing frequency.

\section{Results}
\subsection{Exposure Frequency Affects Adoption (RQ1)}
In this section, we address RQ1 by exploring how the frequency of exposure influences users' likelihood to adopt information sources and how this might vary based on the source's credibility. To this aim, we frame the probability of adoption of an information source as a function of the number of exposures. We examine all user-domain pairs $(u_{t}, d_{i})$, where target user $u_{t}$ is exposed to information source $d_{i}$. For every pair, we calculate $n_{e}$ as the number of exposures prior to $u_{t}$'s adoption of $d_{i}$ to compose a triplet $(u_{t}, d_{i}, n_{e})$. When we do not observe any adoption of $d_{i}$ from $u_{t}$, we use the number of accumulated exposures to $d_{i}$ as $n_{e}$. The probability of adoption at a particular exposure count $N$, indicated as $Pr(A|n_{e}=N)$, is determined as the ratio of user-content pairs $(u_{t}, d_{i})$ resulting in adoption to all pairs with an exposure level of $n_{e}=N$. The distribution of $(u_{t}, d_{i})$ pairs across different exposure frequencies $N$ is detailed in the \textit{Appendix}.

\subsubsection{Adoption likelihood increases with exposure frequency}
\begin{figure}
    \centering
    \begin{subfigure}{.5\columnwidth}
        \centering
        \includegraphics[width=\linewidth]{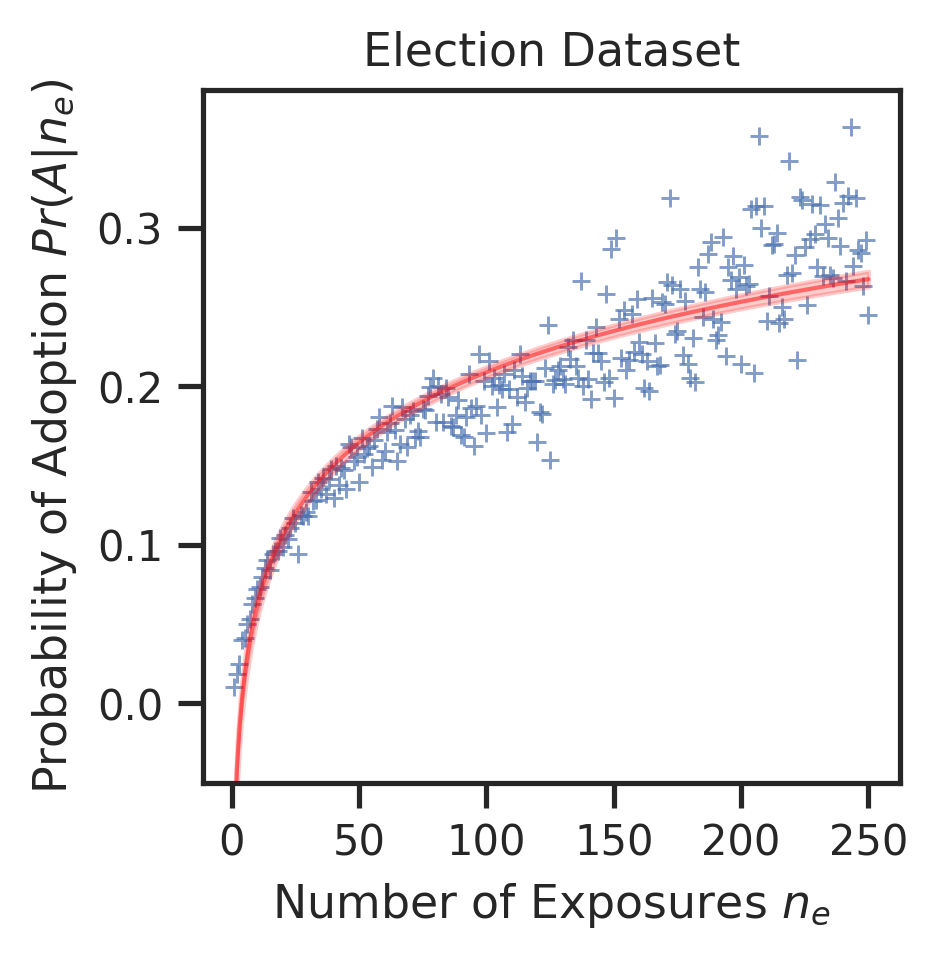}
        \label{prob_adoption_election}
        \vspace{-0.6cm}
    \end{subfigure}%
    \begin{subfigure}{.5\columnwidth}
        \centering
        \includegraphics[width=\linewidth]{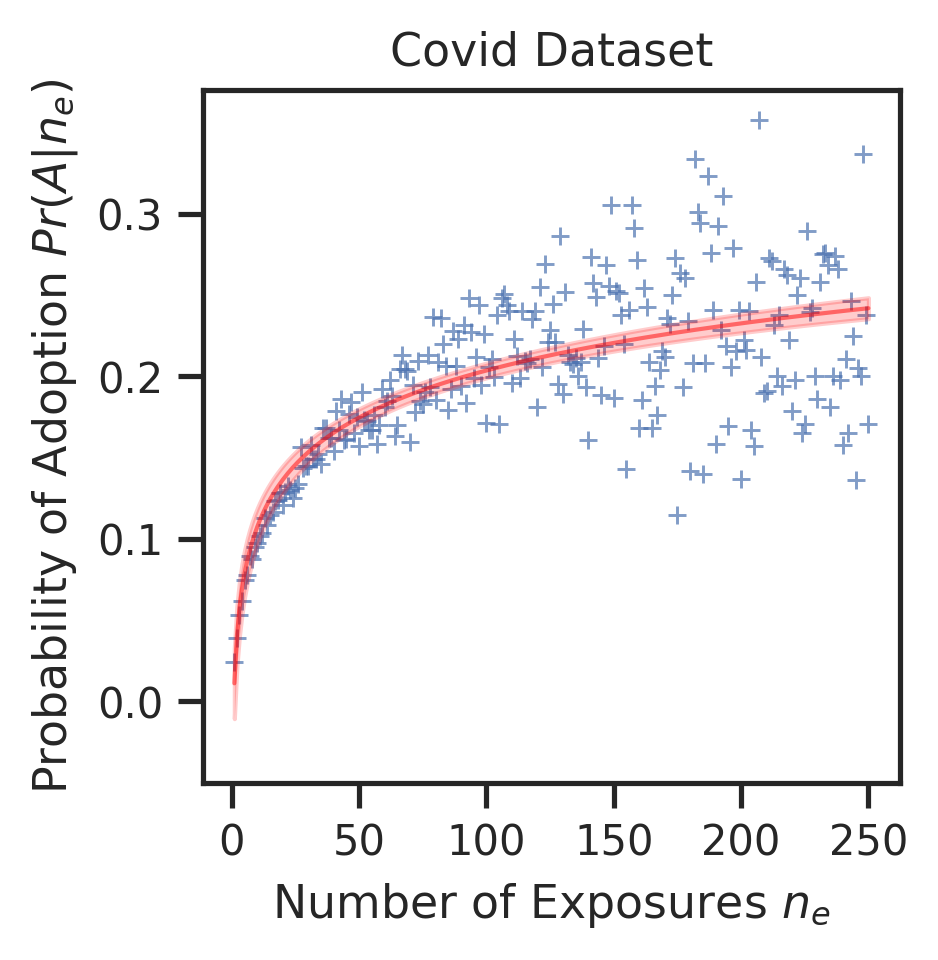}
        \label{prob_adoption_covid}
        \vspace{-0.6cm}
    \end{subfigure}
    \captionsetup{justification=raggedright,singlelinecheck=false}
    \caption{The probability of adoption at varying levels of exposures with a logarithmic regression fit. Each data point represents a specific instance of adoption probability $Pr(A|n_{e})$ at a given exposure $n_{e}$.}
    \label{prob_adoption}
\end{figure}

Our results show that the probability of adoption of information sources  $Pr(A|n_{e})$ increases with the number of exposures $n_{e}$ in both datasets (Mann-Kendall $p$ < .001), as depicted in Figure \ref{prob_adoption}. We set an upper limit of the number of exposures $n_e$ to 250, which encompasses over 99\% of all $(u_{t}, d_{i})$ pairs. This is consistent with similar findings in previous work \cite{hodas2014simple} on the adoption of distinct URLs. Further quantifying this relationship, our linear regression analysis---employing a logarithmic transformation on the number of exposures ($n_e$)---substantiates the influence of exposure on source adoption. Specifically, for the \textsc{Election} dataset, every logarithmic increase in exposure correlates with a 6.4\% rise in the likelihood of adoption ($b$ = .064, $p$ < .001), accounting for 83.2\% of the variability in the data ($R^2$ = .832). Similarly, in the \textsc{Covid} dataset, a log-transformed exposure increment corresponds to a 4.2\% increase in adoption likelihood ($b$ = .042, $p$ < .001), with a model fit explaining 55.4\% of the data variation ($R^2$ = .554). These regression outcomes underscore the role of exposure frequency in users' adoption of information sources.


\subsubsection{Source credibility modulates adoption rate}
\begin{figure}
    \centering
    \begin{subfigure}{.5\columnwidth}
        \centering
        \includegraphics[width=\linewidth]{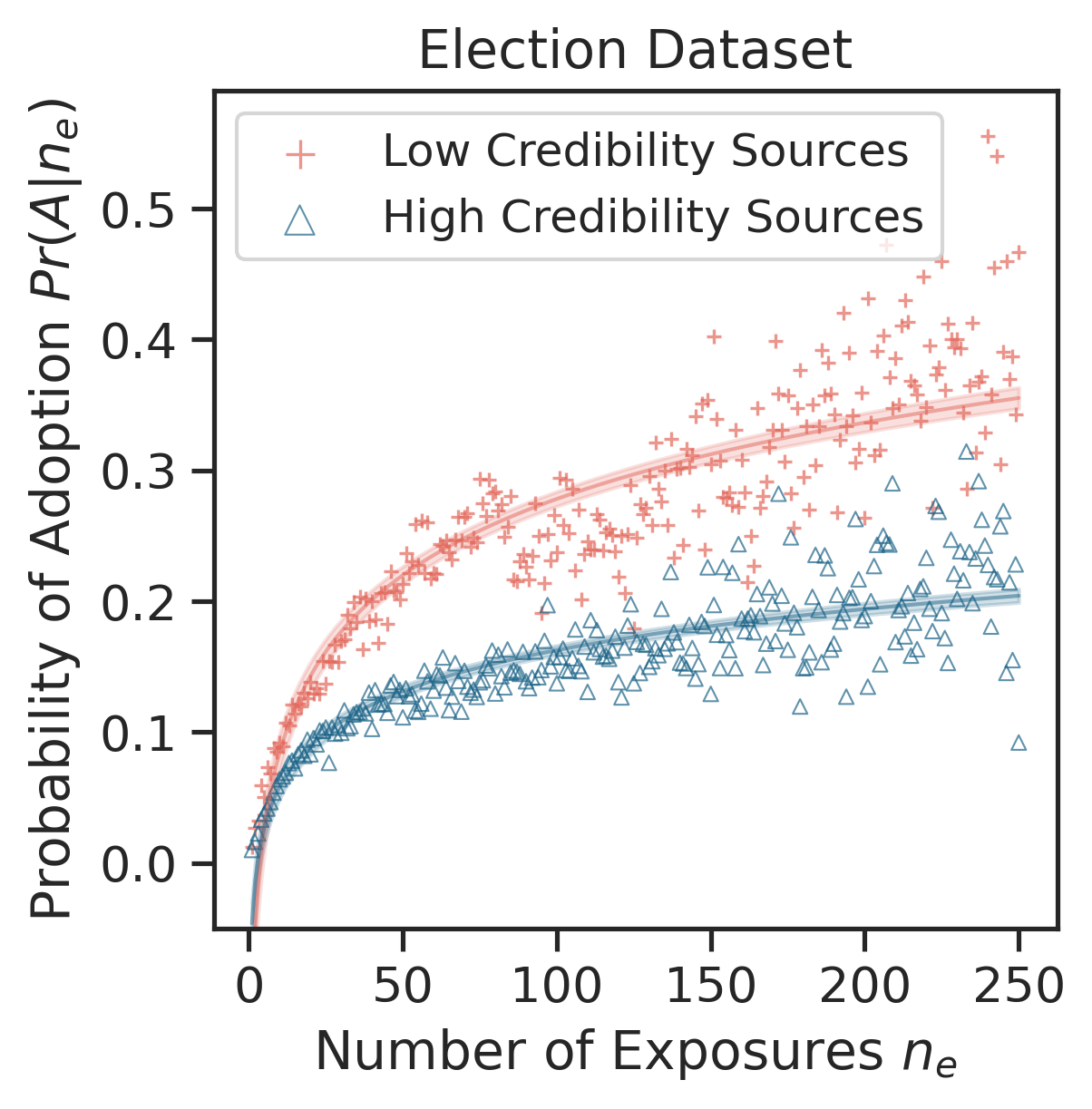}
        \label{prob_content_election}
        \vspace{-0.6cm}
    \end{subfigure}%
    \begin{subfigure}{.5\columnwidth}
        \centering
        \includegraphics[width=\linewidth]{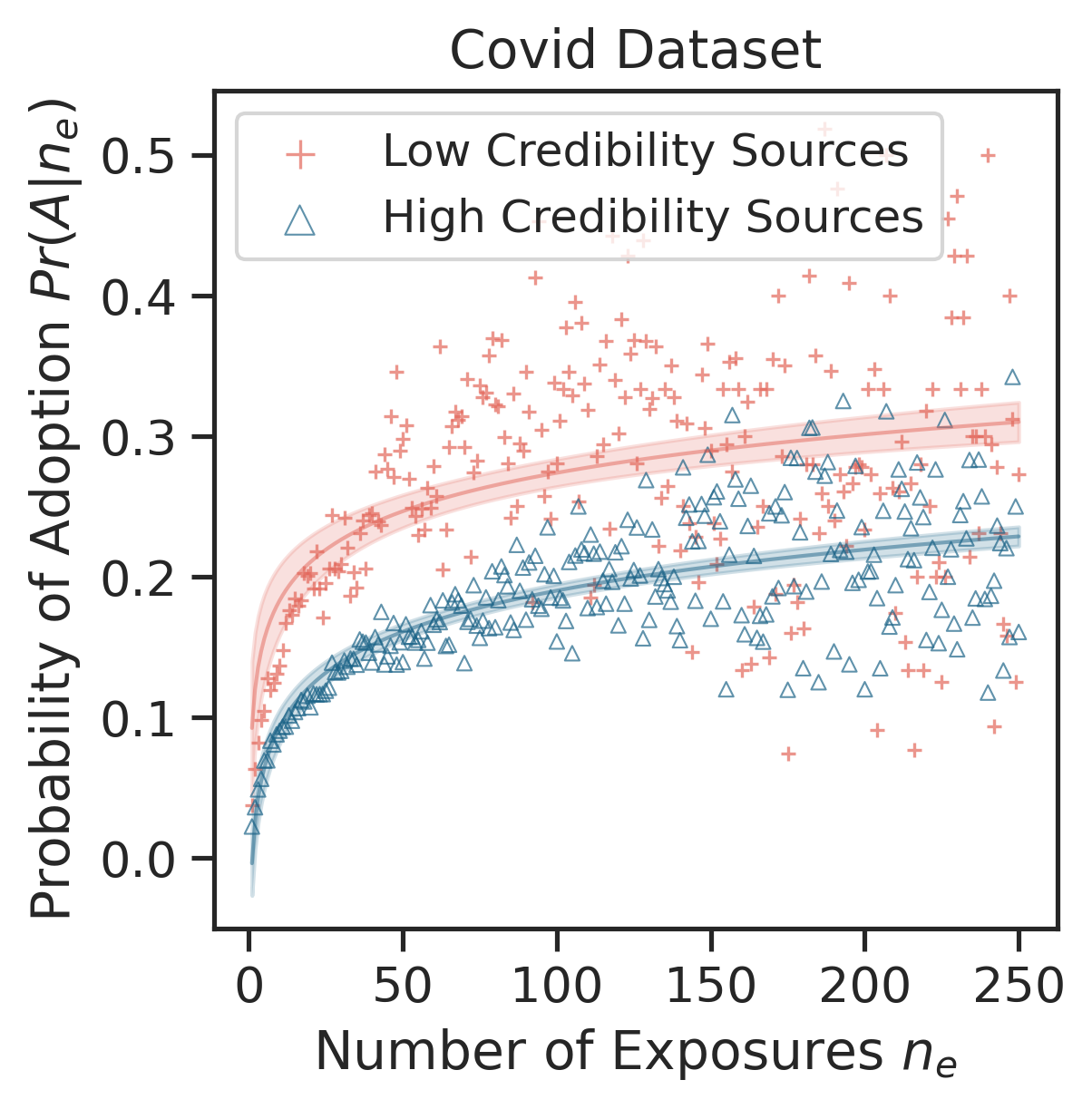}
        \label{prob_content_covid}
        \vspace{-0.6cm}
    \end{subfigure}
    \captionsetup{justification=raggedright,singlelinecheck=false}
    \caption{The probability of adoption at varying levels of exposures with a logarithmic regression fit, modulated by the credibility of information sources.}
    \label{prob_content}
\end{figure}

We now examine the impact of the source's credibility on the relationship between adoption probability and exposure frequency. Figure \ref{prob_content} shows the probability of adoption $Pr(A|n_{e})$ as a function of the number of exposures $n_{e}$ to either high- or low-credibility content. It is evident that, at a specific level of exposure, low-credibility sources exhibit a significantly higher probability of adoption compared to high-credibility sources (Mann-Whitney $p$ < .001). This result is consistent across both datasets.

\begin{figure}
    \centering
    \begin{subfigure}{0.9\columnwidth} 
        \centering        \includegraphics[width=\linewidth]{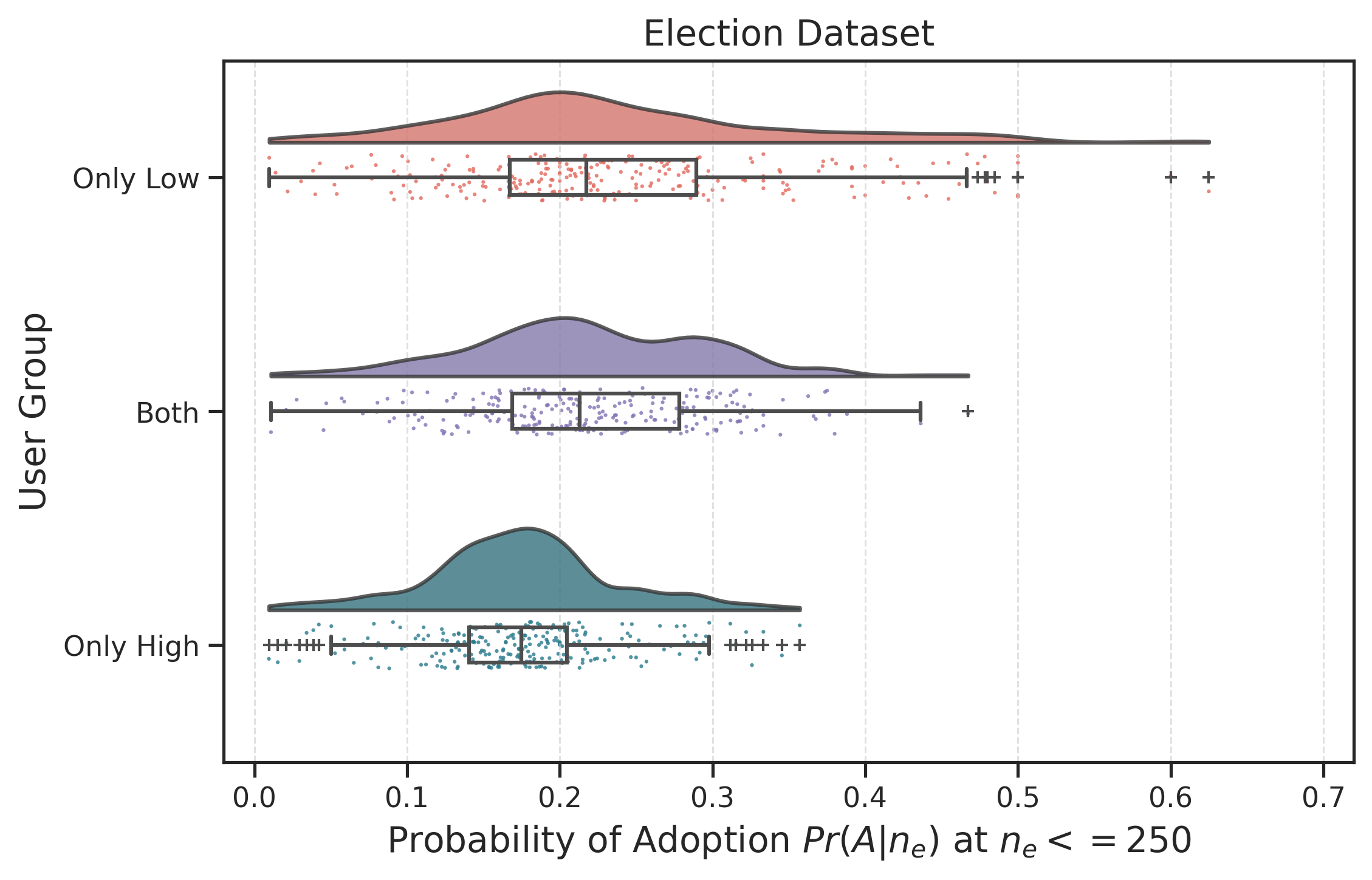}
        \label{prob_user_election}
        \vspace{-0.3cm}
    \end{subfigure}
    
    \begin{subfigure}{0.9\columnwidth} 
        \centering        \includegraphics[width=\linewidth]{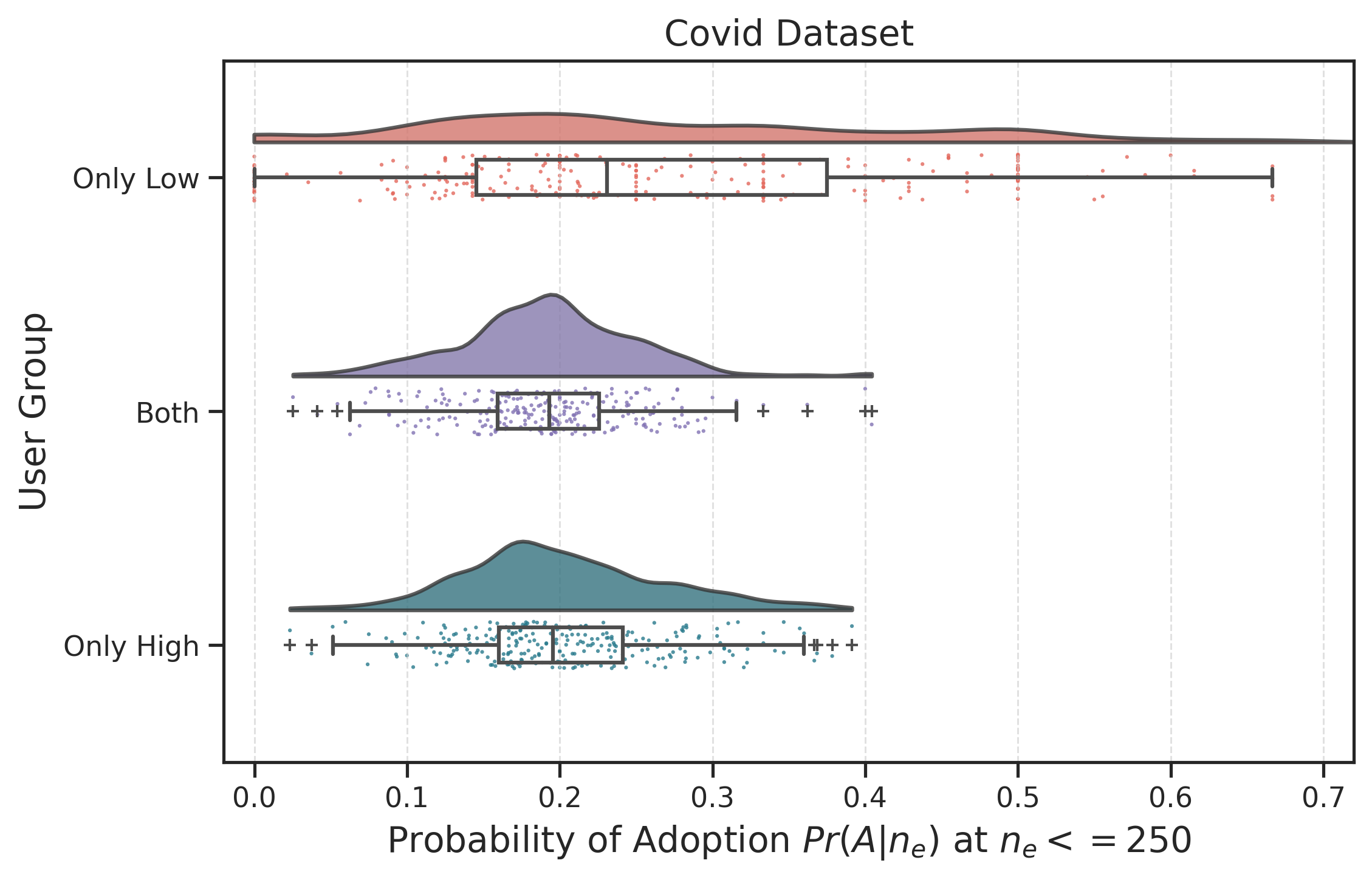}
        \label{prob_user_covid}
        \vspace{-0.5cm}
    \end{subfigure}
    \captionsetup{justification=raggedright,singlelinecheck=false}
    \caption{The probability of adoption given a number of exposures $n_{e}\le250$, modulated by user groups. \vspace{-0.3cm}}
    \label{prob_user_group}
    
\end{figure}

Thus far, we have considered each user-content pair independently, but each user could be exposed to and adopt multiple information sources. Therefore, for further analysis, we distinguish users according to three credibility categories: those sharing exclusively low-credibility sources (``Only Low''), those disseminating only high-credibility content (``Only High''), and those who share both types (``Both''). In Figure \ref{prob_user_group}, we display the probability of adoption given the number of exposures $n_{e}\le250$, taking into account the three distinct groups to which every target user $u_{t}$ belongs.

In the \textsc{Election} dataset, there is no significant difference in the adoption probability between the ``Only Low'' and ``Both'' groups (Mann-Whitney U $p$ = .34). However, the ``Only Low'' group exhibits a higher adoption rate compared to the ``Only High'' group (Mann-Whitney U $p$ < .001). For the \textsc{Covid} dataset, the adoption probability of the ``Only Low'' group is significantly higher than both the ``Only High'' and ``Both'' user groups (Mann-Whitney U $p$ < .001 in both instances). These results underscore that users exclusively adopting low-credibility content consistently exhibit a higher probability of adoption at every exposure level compared to other users. For a more granular breakdown, the reader is directed to 
the \textit{Appendix}.

However, drawing such a conclusion requires careful scrutiny of potential confounding factors. One salient factor is highlighted in a large body of research on online political polarization, which consistently reports that extremely partisan individuals tend to dominate misinformation consumption and dissemination \cite{guess2020exposure, osmundsen2021partisan, jiang2021social, rao2022partisan}. Thus, a crucial question arises: Do our findings remain consistent when excluding users with extremely left or right political leanings?

\begin{table}
\centering
\small
\caption{Comparison of target users' likelihood to adopt low- ($P_{Low}$) and high-credibility ($P_{High}$) sources after removing the top (most right-leaning) and bottom (most left-leaning) x\% of partisan users by political leaning scores.
We apply the Mann-Whitney U test to test that $P_{Low}>P_{High}$ (*** $p<.001$). 
\vspace{-0.2cm}}
\label{user-leaning-check}
\begin{tabular}{lcccccc}
\toprule
 & \multicolumn{3}{c}{\textsc{Election} Dataset} & \multicolumn{3}{c}{\textsc{Covid} Dataset} \\
\cmidrule(lr){2-4} \cmidrule(lr){5-7}
Removal & 10\% & 20\% & 30\% &  10\% & 20\% & 30\% \\
\midrule 
No. of users & 21,038 & 20,361 & 17,886 & 33,921 & 33,460 & 31,803 \\
$\mu_{\text{low}}$ & 0.273 & 0.274 & 0.271 & 0.271 & 0.271 & 0.268 \\
$\mu_{\text{high}}$ & 0.139 & 0.161 & 0.166 & 0.187 & 0.187 & 0.190 \\
 $P_{Low}>P_{High}$ & *** & *** & *** & *** & *** & ***\\
\bottomrule
\vspace{-0.2cm}
\end{tabular}
\end{table}

\subsubsection{Robustness checks with non-partisan users}
To mitigate potential biases that might arise from partisanship, we repeat the same analyses, taking into account news source political partisanship. It is worth noting that a substantial amount of left-leaning sources are rated as credible (84\% with a credibility score $>=$ 0.8 and 93\% $>=$ 0.6). That said, there is still a non-negligible number of non-credible left-leaning sources: 7\% ($n$ = 54) of left-leaning sources have a credibility score $<=0.4$. In contrast, right-leaning sources exhibit a more diverse distribution of credibility scores. About 33\% ($n$ = 478) of these sources have a credibility score of $>=$ 0.6 and 67\% a credibility score of $<=0.4$. This indicates that consumers of both left- and right-leaning news sources have the potential to encounter both credible and non-credible content. 



We repeat our evaluation by filtering out the most partisan users from our target user group. We experiment with retaining users falling within the top and bottom 10\%, 20\%, and 30\% of the political leaning score distribution. 
The results for these varying thresholds, detailed in Table \ref{user-leaning-check}, confirm our previous findings: the probability of adopting low-credibility sources is significantly higher than that of high-credibility sources at varying exposure levels.

\paragraph{Summary:} 
Our findings indicate that \textbf{it takes fewer exposures for individuals to adopt low-credibility sources than highly credible sources}. This suggests that people may be more susceptible to misinformation than factual information when experiencing the same level of exposure, regardless of their political leanings. Expanding on \citet{nikolov2021right}'s insights into the correlation between partisanship and misinformation vulnerability, we provide a more nuanced understanding of how both right- and left-leaning users can be susceptible to misinformation, potentially at different rates, based on their exposure to low-credibility sources.


\subsection{Credibility Modulates the Exposures Needed for Adoption (RQ2)}
While in the previous section, we investigate the probability of adoption based on exposure, in this section, we focus on understanding how many exposures are typically needed to adopt an information source. Specifically, we aim to investigate how the credibility of the source influences the number of exposures leading up to adoption. 
To address RQ2, we assess the frequency of exposures associated with each instance of adoption and then categorize them based on the credibility of the adopted source.

To systematically categorize the relative exposure levels that led to adoption, we aggregate adoption instances across all information sources. Each case of adoption is associated with a specific number of exposures. We subsequently distribute these counts into three equal parts, or terciles, for each dataset: \textit{Low Exposure} (first tercile), \textit{Medium Exposure} (second tercile), and \textit{High Exposure} (third tercile). 


\begin{figure}
\centering
\includegraphics[width=\linewidth]
{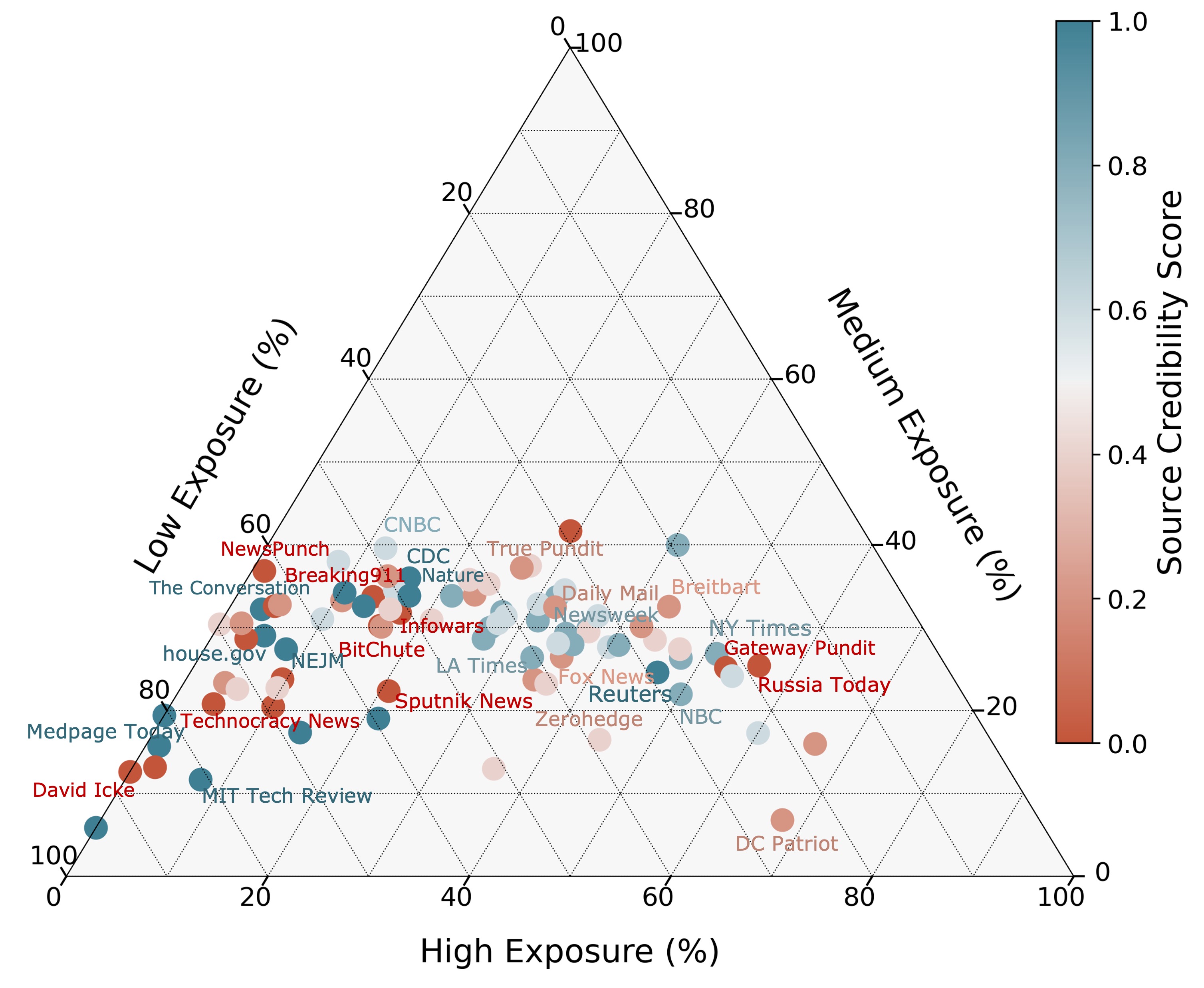}
\caption{Exposure level of the top 15 most frequently adopted information sources at varying credibility for the \textsc{Covid} dataset. Each data point in the simplex is an information source, and the position of the point represents the percentage (\%) of instances where the information source was adopted at low/medium/high exposure levels.}
\label{simplex-covid}
\end{figure}

In Figure \ref{simplex-covid}, we visually represent the distribution of adopted information sources as a function of their exposure levels in the \textsc{Covid} dataset (results are consistent in the \textsc{Election} dataset --- see the \textit{Appendix}). To enhance the clarity of the visualization, we focus on the top 15 most frequently shared information sources across all six credibility levels. For each of these information sources, we depict the percentages of adoption cases at low, medium, and high exposure levels. Each point within the simplex represents an individual information source, with the sum of the percentages always equating to 100\%. The color of each point indicates the credibility of the corresponding information source.
We observe that sources with both very low (dark red points) and very high (dark blue points) credibility require fewer exposures to achieve adoption compared to sources with moderate credibility. Our statistical analysis (Table \ref{pairwise-comparison}), which encompasses all information sources, confirms that sources positioned at the extreme ends of the credibility spectrum are statistically significantly more prone to being adopted with fewer exposures with respect to other sources. Full results of the pairwise comparison tests are available in the \textit{Appendix}.

\begin{table}
\centering
\small
\caption{Comparison of adoption probabilities for varying credibility sources with \textit{Low Exposure} level using one-sided Mann-Whitney U tests (* $p < .05$, ** $p < .01$, *** $p < .001$, N.S.= not significant). Extremes of the credibility spectrum (\textit{Very Low} and \textit{Very High}) are compared with moderate credibility sources (\textit{Low}, \textit{Mixed}, \textit{Mostly Factual}, \textit{High}).
\vspace{-0.4cm}
} 
\label{pairwise-comparison}
\begin{tabular}{@{}lrcccc@{}}
\toprule
& & $P_{\textit{Low}}$ & $P_{\textit{Mixed}}$ & $P_{\textit{Mostly Factual}}$ &  $P_{\textit{High}}$ \\
\midrule 
\textsc{Election}  & $P_{\textit{Very Low}} $ & >** & >* & >*** & >*** \\
Dataset & $P_{\textit{Very High}} $ & >*** & >** & >*** & >*** \\
\midrule
\textsc{Covid} & $P_{\textit{Very Low}}$ & N.S. & N.S. & >** & >** \\
Dataset & $P_{\textit{Very High}}$& >** & >*** & >*** & >***\\
\bottomrule
\end{tabular}
\end{table}



\paragraph{Summary:} We show that \textbf{both extremely low and high credibility sources require fewer exposures for adoption}. This swift adoption may be indicative of polarization in user behavior. For extremely low credibility sources, users might be drawn to these due to their novel or controversial nature \cite{vosoughi2018spread}, the allure of ``alternative'' news \cite{stocking2022role}, or because they resonate with fringe beliefs \cite{bauer2021believing}. For extremely high credibility sources, users might be seeking authoritative information, looking for a trusted and definitive viewpoint on a topic \cite{yaqub2020effects}. This could also reflect a segment of the audience that highly values accuracy and reliability. The fact that both extremes require fewer exposures to be adopted suggests that users may have strong emotional or cognitive reactions to these types of sources, leading to quicker decisions about adoption \cite{baum2021emotional}.

\subsection{Adoption Credibility Is Driven by Exposure Credibility (RQ3)}
In addressing RQ3, we present an analysis to gain a deeper understanding of individual user behavior concerning the sources they encounter and choose to adopt. Our objective is to investigate whether exposure to a particular set of sources interplays with users' decision to adopt sources with either similar or dissimilar levels of credibility. 

\subsubsection{Exposure credibility correlates with adoption credibility}
\begin{figure}
\centering
\includegraphics[width=6.5cm]
{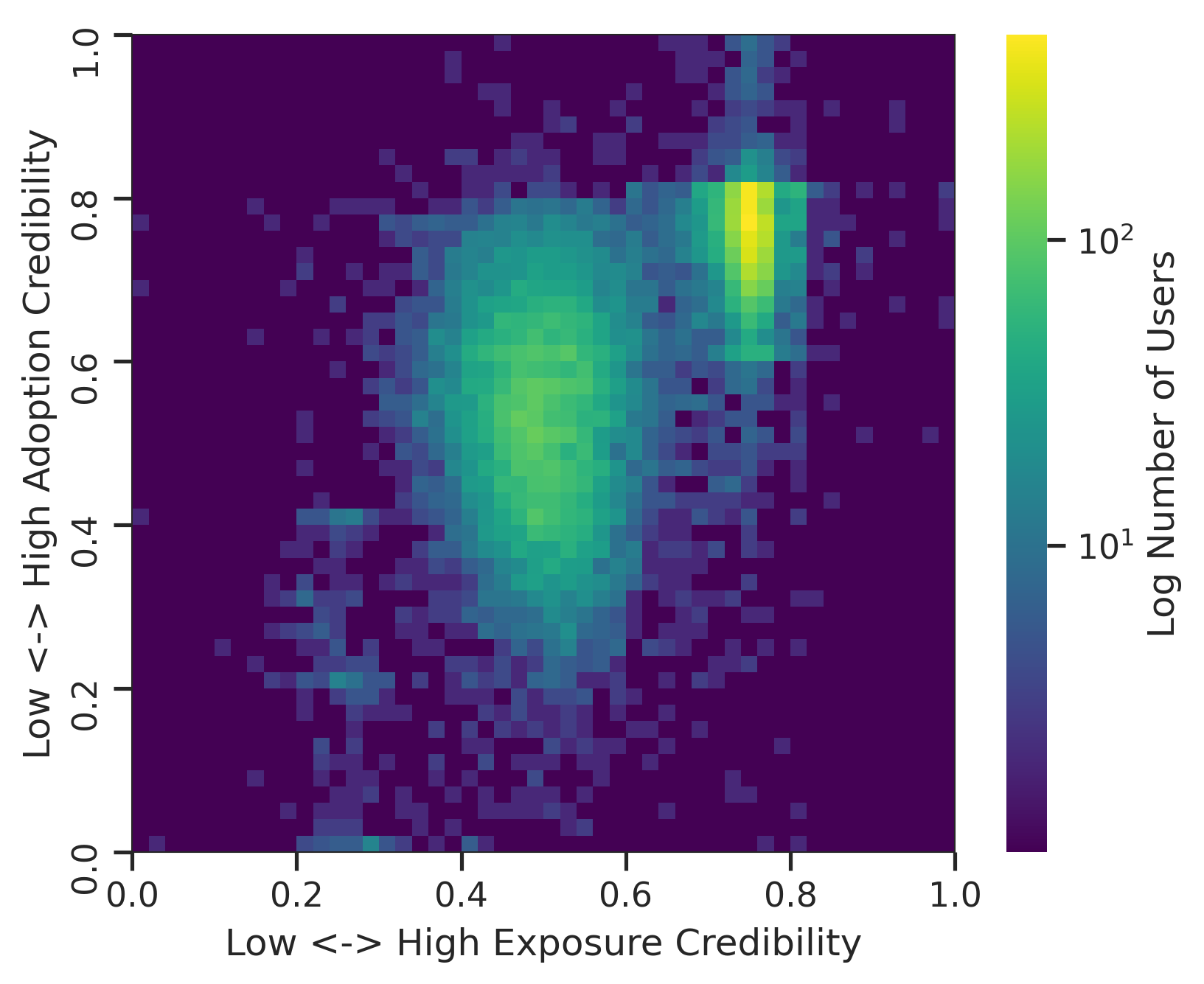}
\caption{Correlation between exposure and adoption credibility scores for the \textsc{Election} dataset.\vspace{-0.2cm}}
\label{heatmap}
\end{figure}

To begin, we examine the link between the credibility of sources users are exposed to (\textit{exposure credibility}) and the credibility of sources they adopt (\textit{adoption credibility}) in Figure \ref{heatmap}. Notably, there is a prominent concentration of users on the diagonal, as well as within the upper- and mid-range of both exposure and adoption credibility scores. This suggests that while the majority of users predominantly encounter and share highly credible sources, a substantial portion of users are also susceptible to less credible information. Additionally, we observe a strong positive Pearson's correlation between exposure and adoption credibility scores for both datasets, with $r$ = 0.64 ($p$ < .001) for the \textsc{Election} dataset and $r$ = 0.66 ($p$ < .001) for the \textsc{Covid} dataset.

\subsubsection{Same-credibility exposure precedes same-credibility adoption} \label{rq3}
\begin{figure}
\centering
\includegraphics[width=7.5cm]
{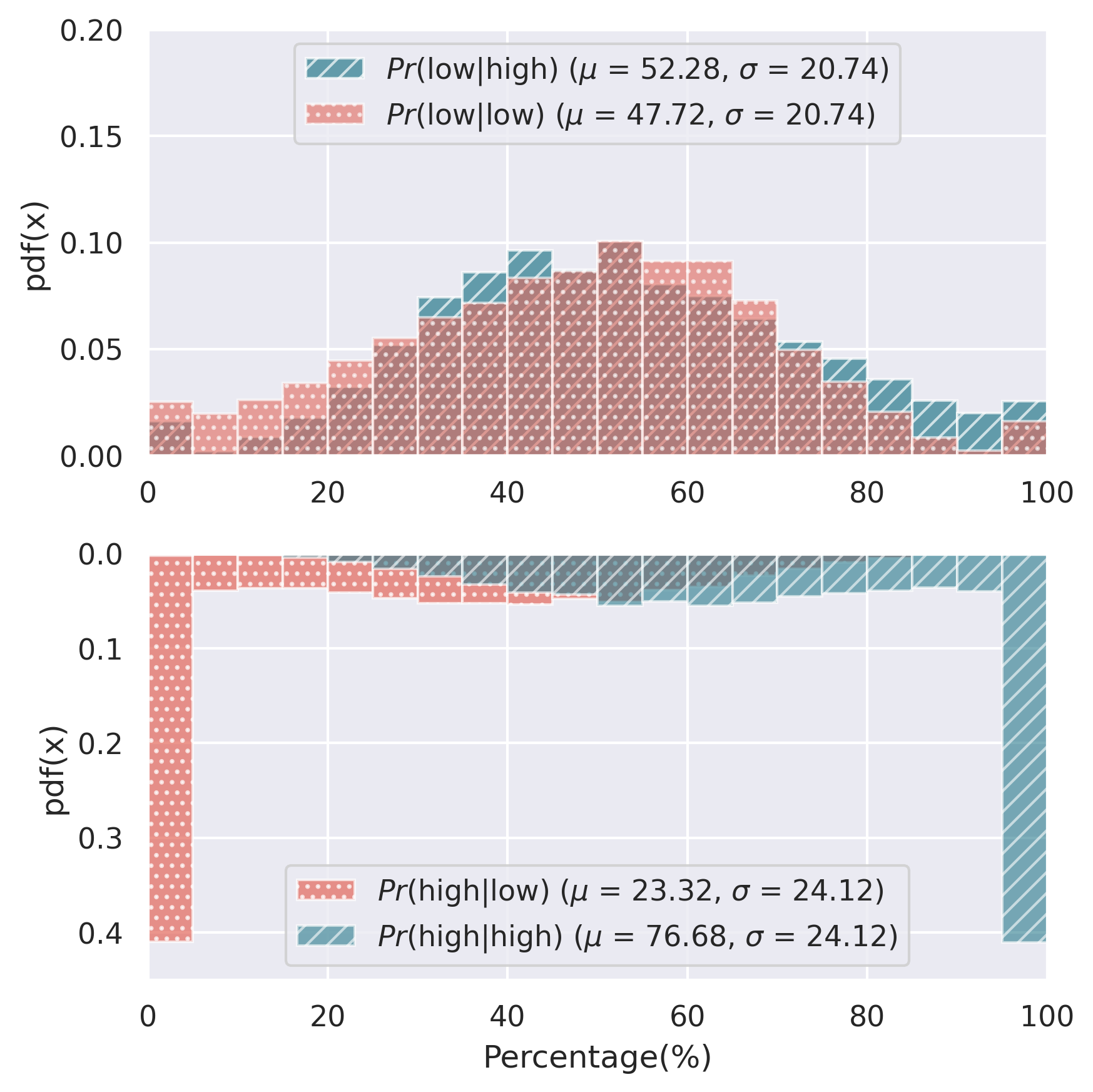}
\caption{Distribution of the percentages of low- and high-credibility exposures in the 7 days prior to low- and high-credibility adoptions in the (\textsc{Election} dataset). $Pr(high|low)$, for example, refers to the probability of high-credibility adoption given the percentage of low-credibility exposure. 
}
\label{bayesian-plot}
\end{figure}

We extend our analysis to investigate the likelihood of users' adoption of low- and high-credibility sources conditional on the credibility of their prior exposures. To achieve this, we observe all the posts that a target user $u_{t}$ might have been exposed to within a 7-day period prior to adopting an information source. Subsequently, we calculate the proportion of these prior exposures originating from low- and high-credibility sources. The resulting distribution of these prior exposures is visualized in Figure \ref{bayesian-plot}. 

The two quadrants depict the probability distributions for adopting low-credibility (top panel) and high-credibility (bottom panel) sources, depending on the percentage of exposure to low-credibility (pink distribution) or high-credibility (light blue distribution) sources in the \textsc{Election} dataset.
Results indicate that, prior to posting a tweet containing a low-credibility source, users are exposed to an average of 48\% ($\pm$0.21) low-credibility sources and 52\% ($\pm$0.21) high-credibility sources.
In contrast, prior to posting a tweet containing a high-credibility source, users are exposed to an average of 77\% (±0.24) high-credibility sources, which is 25\% higher than in low-credibility adoptions.
To prove the statistical significance of these differences, we run a Mann–Whitney U test comparing the probabilities of exposure to high- versus low-credibility sources prior to low-credibility adoptions ($Pr(high|low)$ and $Pr(low|low)$), and between high- and low-credibility sources prior to high-credibility adoptions ($Pr(high|high)$ and $Pr(low|high)$), with both comparisons yielding significant results ($p$ < .001).

It is important to highlight that the probability $Pr(low|low)$ is significantly lower than $Pr(high|high)$ (Mann-Whitney U $p$ < .001). This observation suggests that individuals who share high-credibility content tend to require a notably higher proportion of high-credibility exposure before adopting such sources, likely because they are less susceptible to content exposure. In contrast, those who adopt low-credibility sources may adopt it with approximately 50\% exposure to low-credibility sources. As a robustness check, we experiment with different prior exposure windows (three days and one day) and conduct a parallel analysis on the \textsc{Covid} dataset,  detailed in the \textit{Appendix}. The results from these additional analyses closely align with those obtained in the primary analysis.

\paragraph{Summary:} Our findings reveal that \textbf{the overall credibility of sources a user is exposed to aligns with the overall credibility of sources the user eventually adopts}. 
Although this pattern holds true for users who adopt both high- and low-credibility sources, it is important to note that the former group appears to exhibit lower trust levels, necessitating more exposure compared to the latter group. This analysis reinforces the presence of echo chambers as previously highlighted in the literature \cite{rao2022retweets,sasahara2021social}. However, our study extends this understanding by exploring the impact of source credibility, suggesting that echo chambers on Twitter may not only be self-selective based on political ideology but also influenced by the credibility of the content adopted by users.

\section{Discussion and Conclusions}
This study, grounded in millions of tweets, delves into the intricate dynamics of information source adoption on Twitter, with a particular emphasis on the pivotal roles played by source credibility and user exposure. We present three main insights from our research.

First, we show that users' susceptibility to influence is associated with the credibility of the information to which they are exposed. Adoption of low-credibility sources tends to occur with fewer prior exposures than the adoption of high-credibility sources, a trend we find consistent at the individual user-source level and aggregated at the user level. Though source credibility is correlated with partisanship, with most credible sources being left-leaning, we find that source partisanship alone does not explain the exposure variability in high- vs. low-credibility source adoption.

Second, when considering all adopted sources, both extremely low- and high-credibility sources require less exposure before adoption, whereas moderately credible sources need more. One theory to explain this dichotomy of information adoption is that low-credibility sources are prone to be adopted by users who are easily susceptible, whereas high-credibility sources are prone to be adopted by those who trust the source as official or authoritative. Further research could take into account media source popularity, sensational language use, user demographics, and cognitive biases for a more in-depth exploration. 

Finally, we show the credibility of sources from past exposure aligns with the credibility of sources the users adopt, emphasizing how users' prior experience can reinforce and possibly steer their future choices and behavior. Our research suggests that misinformation susceptibility could be linked to a tendency to trust content too swiftly. As such, vulnerability to misinformation among susceptible users could be exacerbated by the frequent circulation of such content within their networks. 

Beyond these insights, our study contributes methodologically by operationalizing the theoretical concept of susceptibility as a probability function of exposure and adoption, independent of content credibility. We thereby offer a more nuanced view of how individuals interact with information sources, whether credible or not. Future research could extend our analysis by constructing a generalized framework to assess users' susceptibility to \textit{any} piece of information, potentially taking into account network structures or the psychological profiles of users. 

\paragraph{Implications.} Our research provides several practical implications for social media platforms and policymakers. The observed correlation between the frequency of exposure to certain credibility levels of information and subsequent adoption underscores the potential of modifying platform algorithms to diminish the spread of low-credibility content (e.g., shadowbanning \cite{jaidka2023silenced}). Since the credibility of users' adoption tends to align with the credibility level of their most common exposures, platforms can design interventions to promote high-credibility sources (e.g., crowdsourced fact-checking \cite{pennycook2020fighting}). Additionally, our research lends support to the implementation of countermeasures against misinformation, such as increased transparency about source credibility (e.g., displaying news source trustworthiness ratings \cite{celadin2023displaying}) and the proactive dissemination of factual information to users frequently exposed to misinformation.

\paragraph{Limitations.}
We recognize certain limitations of our research. First, our study relies on datasets centered around two specific events: the 2020 US election and the COVID-19 pandemic. Therefore, generalizing our results to different contexts or other social media platforms requires additional validation. Second, we infer exposure from observable user interactions, which may not reflect an accurate representation of their actual exposures. However, given the constraints of our dataset, this serves as a practical proxy for accessing exposure, consistent with previous work \cite{rao2022partisan, ferrara2015measuring, rao2022retweets}. Third, we consider engagements with an information source as an adoption, which can potentially be an indicator of endorsement,
but this may not always be the case. Replies and quotes, for instance, may signify disagreement or critique rather than support \cite{hemsley2018tweeting}. Despite these caveats, the robustness of our results across multiple tests lends credibility to our research outcomes.

\paragraph{Ethical considerations.}
Our study's exploration into the effects of source credibility and exposure on information adoption on Twitter bears the dual-edged potential of being misapplied, particularly in engineering information cascades that could amplify misinformation. We are aware of these risks, yet posit that the insights gleaned offer substantial value in informing strategies to combat the spread of unreliable information online. This work is conducted with a commitment to ethical standards and has been reviewed and approved by our institution's IRB.

\begin{acks}
Work supported in part by DARPA (contract \#HR001121C0169).
\end{acks}

\bibliographystyle{ACM-Reference-Format}
\bibliography{sample-base}


\section*{Appendix}

\subsection{Information Source Details}
Table \ref{info_source} displays the most frequently shared information sources across varying credibility levels. 
\subsection{Exposure Frequency and Probability of Adoption}
\begin{figure}
    \centering
    \begin{subfigure}{.5\columnwidth}
        \centering
        \includegraphics[width=\linewidth]{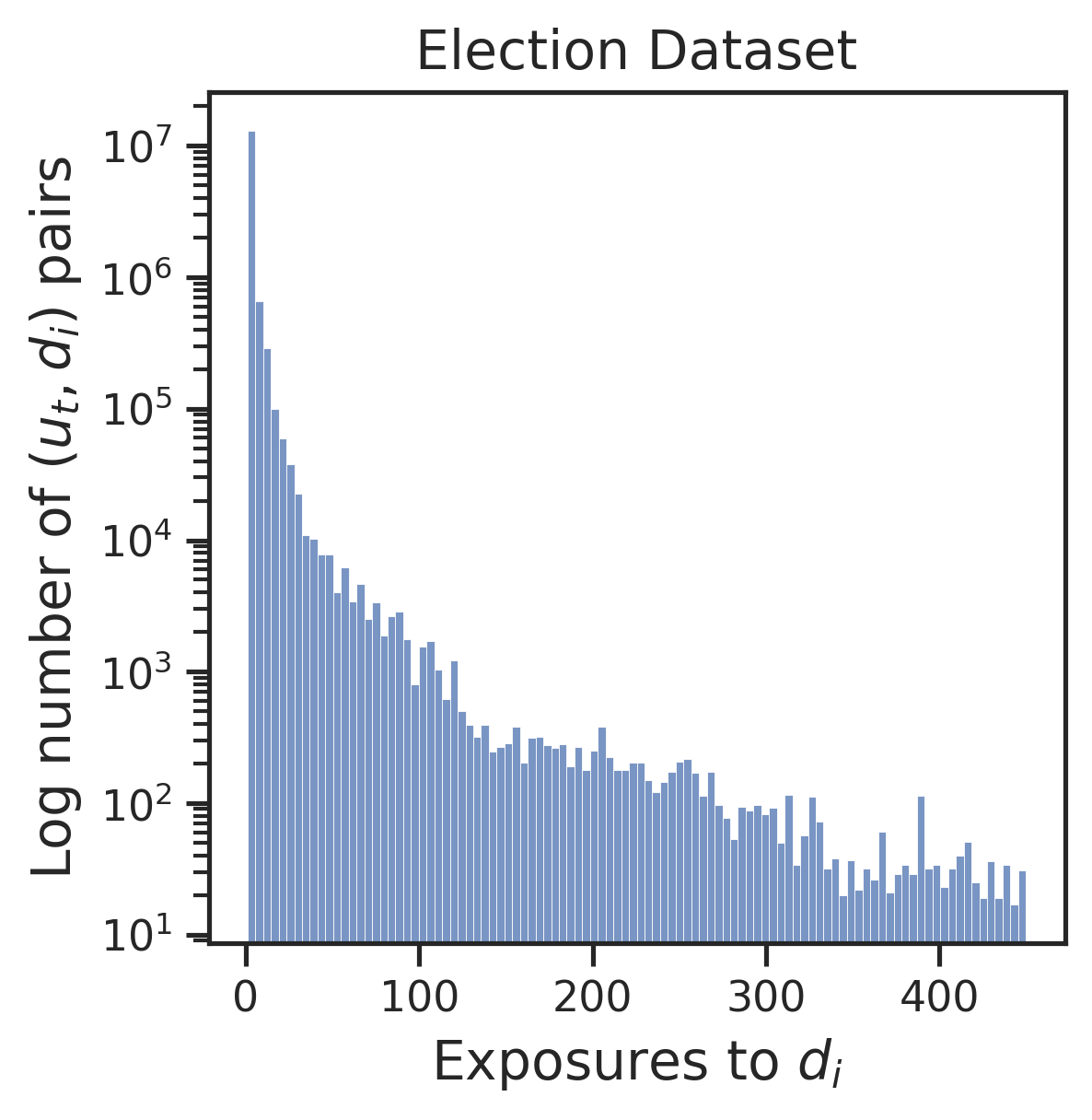}
        \caption{\textsc{Election} dataset}
        \label{user-content-pairs_election}
    \end{subfigure}%
    \begin{subfigure}{.5\columnwidth}
        \centering
        \includegraphics[width=\linewidth]{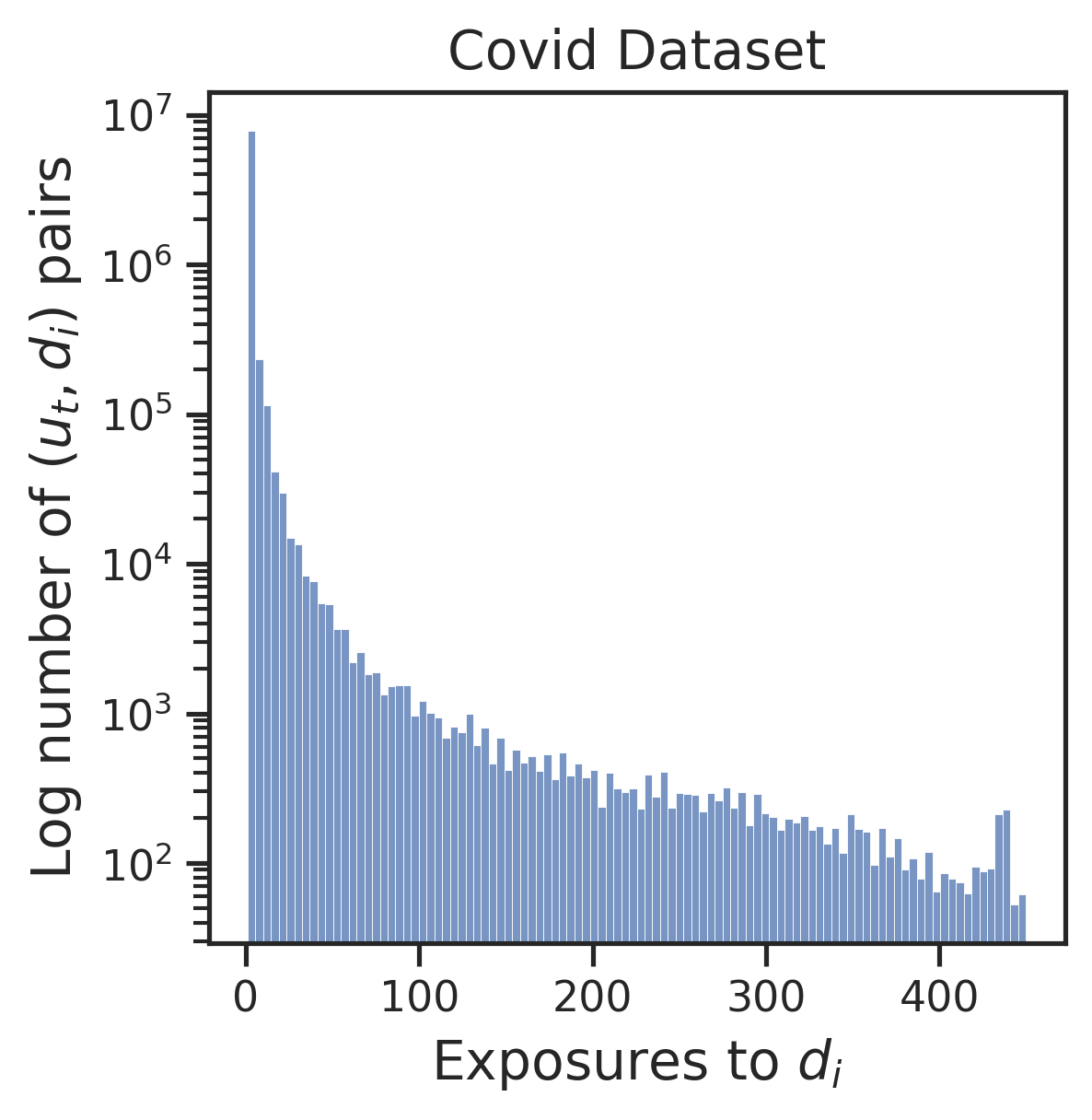}
        \caption{\textsc{Covid} dataset}
        \label{user-content-pairs_covid}
    \end{subfigure}
    \captionsetup{justification=raggedright,singlelinecheck=false}
    \caption{Distribution of the exposure count.}
    \label{user-content-pairs}
\end{figure}

\begin{figure}[H]
    \centering
    \begin{subfigure}{0.7\linewidth}
        \centering
        \includegraphics[width=\linewidth]{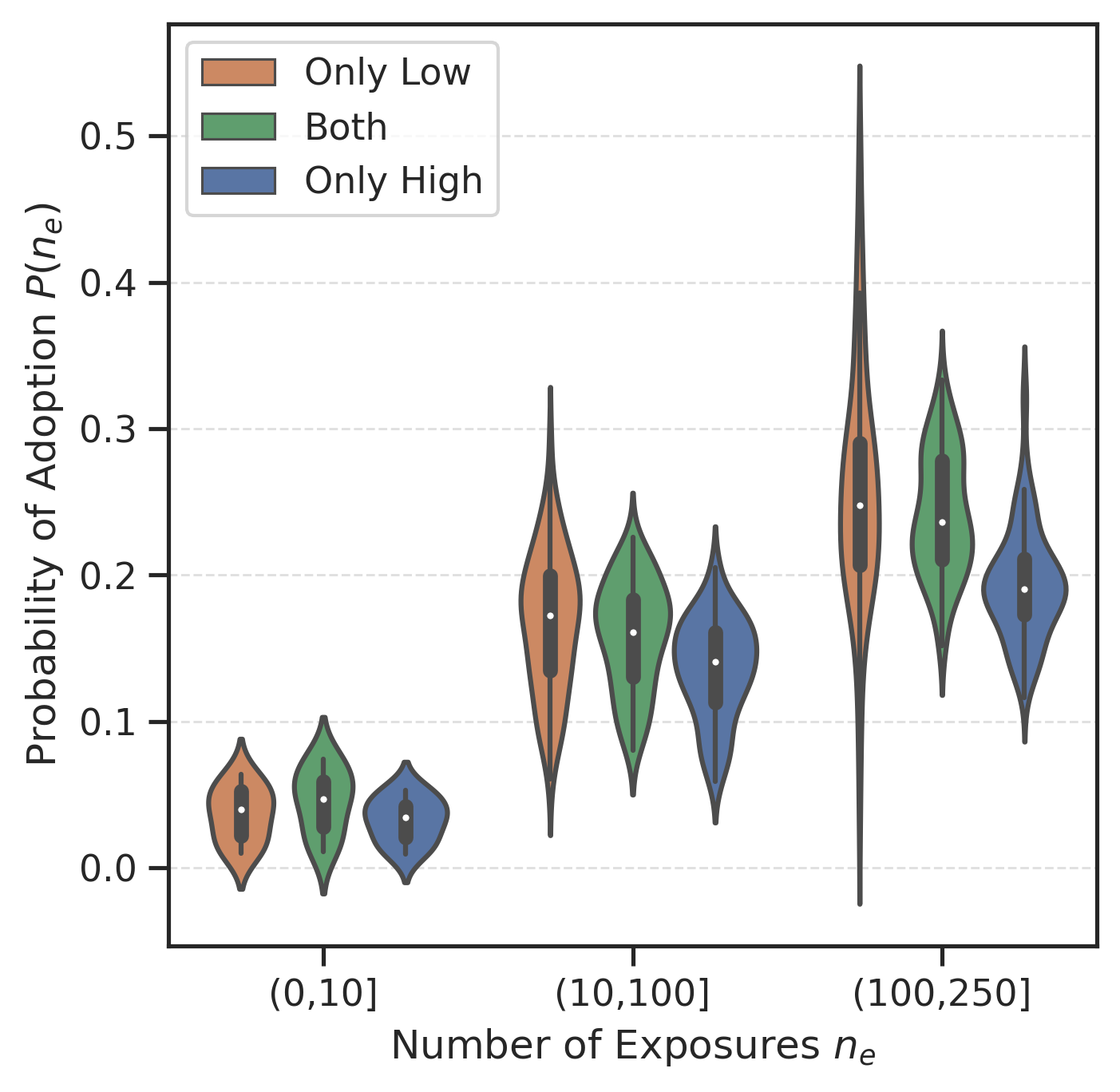}
        \caption{\textsc{Election} dataset}
        \label{prob_adoption_user_election}
    \end{subfigure}%
    \quad
    \begin{subfigure}{0.7\linewidth}
        \centering
        \includegraphics[width=\linewidth]{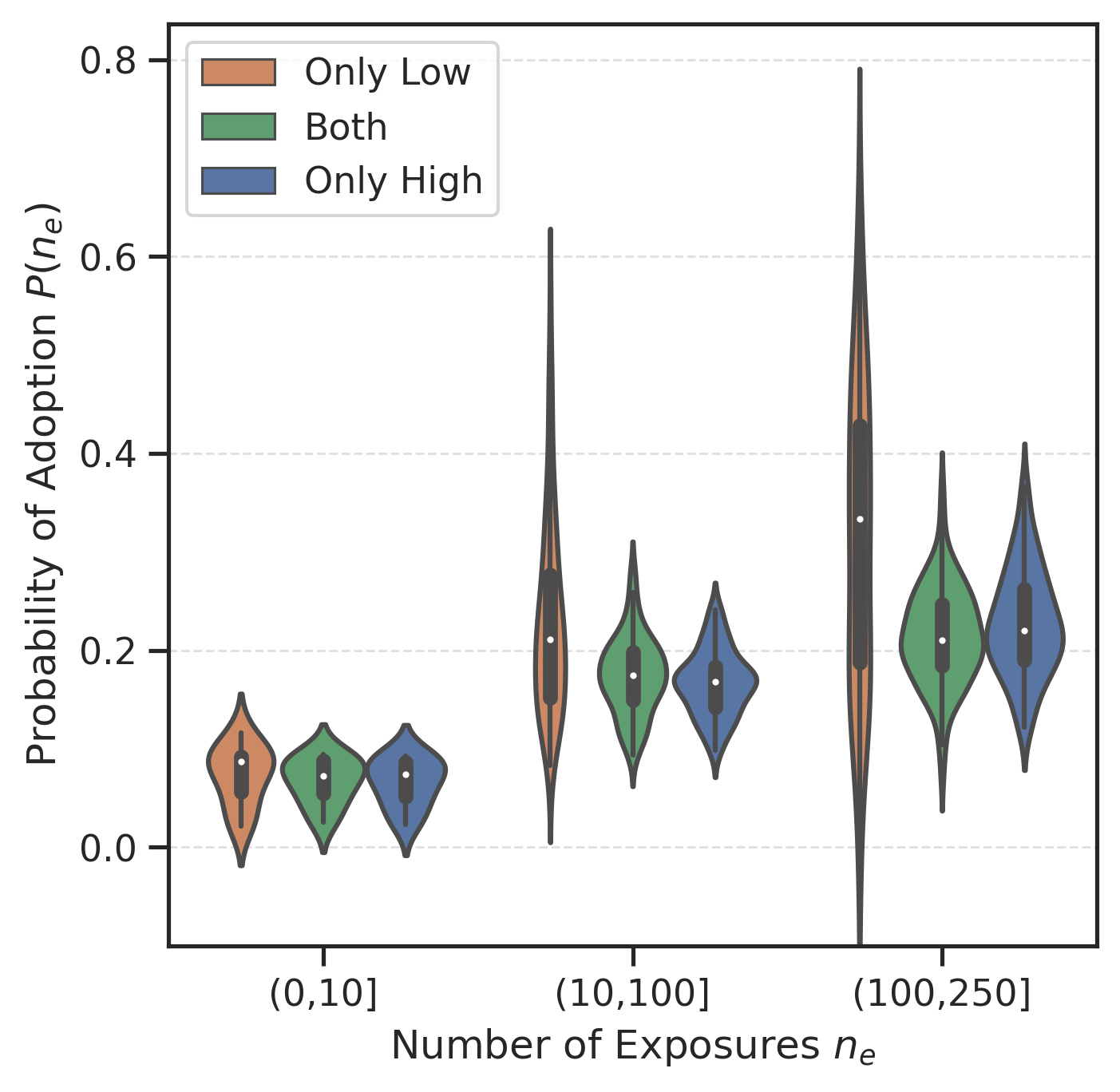}
        \caption{\textsc{Covid} dataset}
        \label{prob_adoption_user_covid}
    \end{subfigure}
    \captionsetup{justification=raggedright,singlelinecheck=false}
    \caption{The probability of adoption at varying levels of exposures modulated by user groups. 
    }
    \label{prob_adoption_user_group}
\end{figure}

Figure \ref{user-content-pairs} shows the distribution of $(u_{t}, d_{i})$ pairs across different exposure frequency $N$. Both histograms exhibit characteristics of a \textit{heavy-tail} distribution, where a small number of $(u_{t}, d_{i})$ pairs have a very large number of exposures, while the majority have very few. 
Figure \ref{prob_adoption_user_group} depicts the probability of adoption as a function of exposure frequency, modulated by the three user groups: those sharing exclusively low-credibility sources (``Only Low''), those disseminating only high-credibility content (``Only High''), and those who share both types (``Both''). To provide a finer-grained analysis of exposure, we categorize the exposure frequency into three bins: (0,10], (10,100], and (100,250].

\renewcommand\floatpagefraction{.1}
\begin{table}
  \centering
  \small
\caption{The top 5 most commonly shared information sources within each credibility level in both the \textsc{Election} and \textsc{Covid} datasets.}
\label{info_source}
  \begin{tabular}{@{}lp{1.3cm}rr@{}}
  \toprule
    \multicolumn{2}{l}{\textbf{Credibility}} & \textbf{\textsc{Election} Dataset} & \textbf{\textsc{Covid} Dataset} \\
    \midrule
    \multicolumn{3}{l}{\textit{Low-Credibility Sources}}\\
    \midrule
    & Very Low & rt.com & rt.com  \\
    & ($N$ = 197) & thegatewaypundit.com & thegatewaypundit.com \\
    & & bitchute.com & bitchute.com \\
    & & newspunch.com & breaking911.com \\
    & & technocracy.news & infowars.com \\
    \midrule
    & Low  & davidharrisjr.com & dailymail.co.uk \\
    &($N$ = 470) & theconservativetreehouse.com & zerohedge.com \\
    & & donaldjtrump.com & truepundit.com \\
    & & thedcpatriot.com & swarajyamag.com \\
    & & hannity.com & oann.com \\
    \midrule
    & Mixed & breitbart.com & foxnews.com \\
    &($N$ = 813)  & foxnews.com & breitbart.com \\
    & & thefederalist.com & justthenews.com \\
    & & washingtontimes.com & washingtontimes.com \\
    & & justthenews.com & theepochtimes.com \\
    \midrule
    \multicolumn{3}{l}{\textit{High-Credibility Sources}} \\
    \midrule
    & Mostly  & whitehouse.gov & washingtonpost.com\\
    &Factual  & washingtonpost.com & cnbc.com\\
    &($N$ = 164) & wsj.com & thehill.com \\
    & & thehill.com & today.com \\
    & & nationalreview.com & wsj.com\\
    \midrule
    & High& caller.com & nytimes.com\\
    & ($N$ = 1,700)  & cbsnews.com & nbcnews.com \\
    & & nytimes.com & cbsnews.com \\
    & & ew.com & al.com \\
    & & politico.com & politico.com \\
    \midrule 
    & Very High& house.gov & reuters.com\\
    & ($N$ = 77)  & reuters.com & cdc.gov \\
    & & cdc.gov & nature.com \\
    & & gallup.com & house.gov \\
    & & factcheck.org & bmj.com \\
    \bottomrule
\end{tabular}
\end{table}

\begin{table}[H]
\centering
\small
\caption{Pairwise comparison of adoption probabilities for varying credibility sources at the \textit{High Exposure} level. Mann-Whitney U test *$p < .05$, **$p < .01$, ***$p < .001$, N.S. = not significant. Extremes of the credibility spectrum (\textit{Very Low} and \textit{Very High}) are compared with moderate credibility sources (\textit{Low}, \textit{Mixed}, \textit{Mostly Factual}, \textit{High}).} 
\label{pairwise-comparison-all}

\begin{tabular}{@{}lrcccc@{}}
\toprule
& & $P_{\textit{Low}}$ & $P_{\textit{Mixed}}$ & $P_{\textit{Mostly Factual}}$ &  $P_{\textit{High}}$ \\
\midrule 
\textsc{Election}  & $P_{\textit{Very Low}} $ & <** & N.S. & <** & <** \\
Dataset & $P_{\textit{Very High}} $ & <** & <* & <*** & <***  \\
\midrule
\textsc{Covid} & $P_{\textit{Very Low}}$ &N.S. & <* & <* & <** \\
Dataset & $P_{\textit{Very High}}$&<* & <** & <** & <*** \\
\bottomrule
\end{tabular}
\end{table}

\subsection{Exposures Needed for Adoption}
\begin{figure}[H]
\centering
\includegraphics[width=\linewidth]
{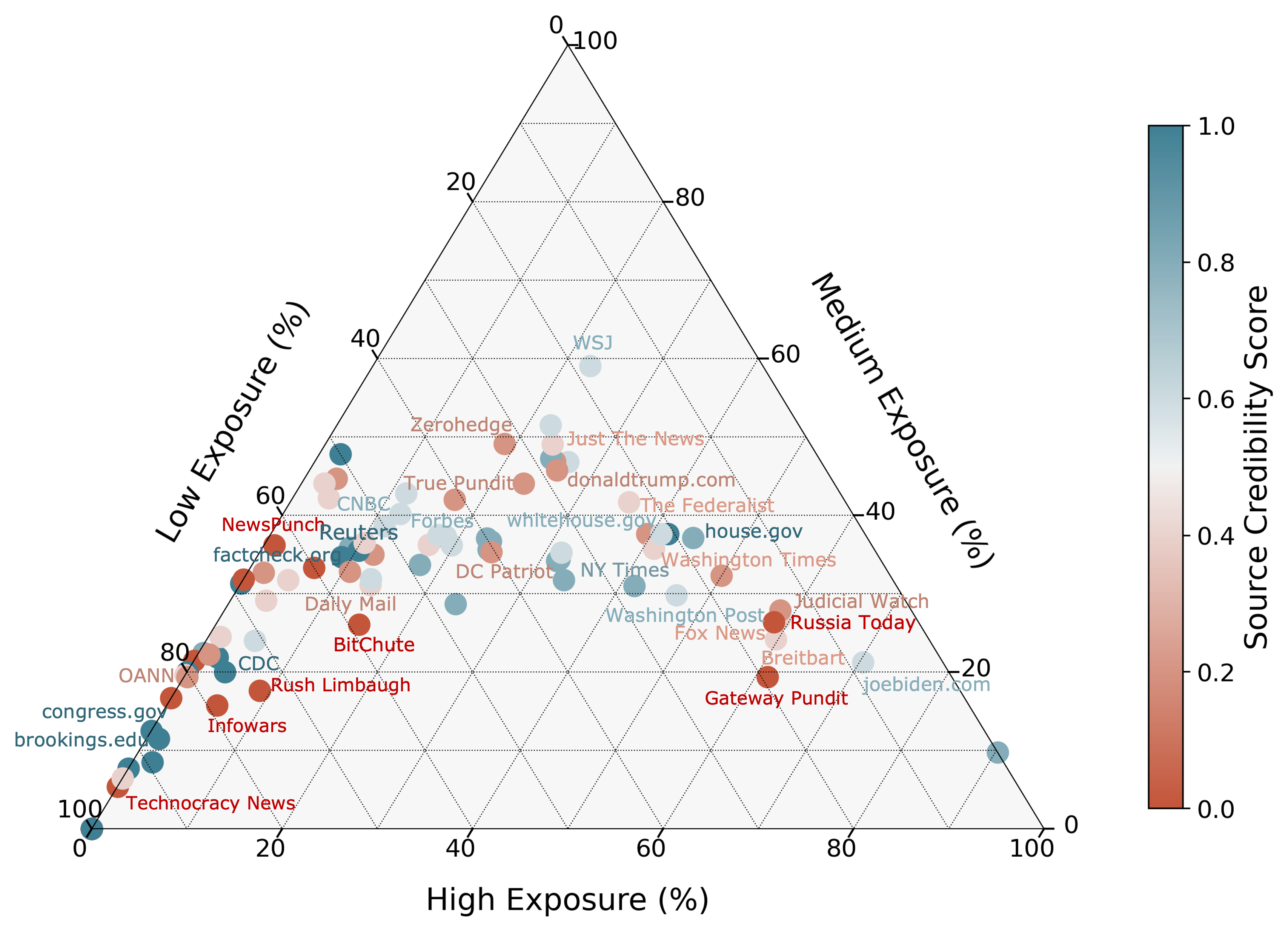}
\caption{Visualization of the exposure amount of the top 15 most frequently adopted information sources from every credibility level for the \textsc{Election} dataset, similar to Figure \ref{simplex-covid}.}
\label{simplex-election}
\end{figure}

In Figure \ref{simplex-election}, we visually represent the distribution of adopted information sources as a function of their exposure levels in the \textsc{Election} dataset. Similar to Figure \ref{simplex-covid}, we display the top 15 most frequently shared information sources across all six credibility levels. Table \ref{pairwise-comparison-all} presents comprehensive results of pairwise comparison tests conducted between sources of differing credibility levels, considering adoption at \textit{Low} and \textit{High Exposure} levels, respectively. Our findings indicate that sources positioned at the extremes of the credibility spectrum are more frequently adopted at the \textit{Low Exposure} level, less frequently at the \textit{High Exposure} level, with no discernible differences at the \textit{Medium Exposure} level.

\subsection{Low-/High-Credibility Exposure Prior to Low-/High-Credibility Adoption}
\begin{figure}[H]
\centering
\includegraphics[width=7cm]
{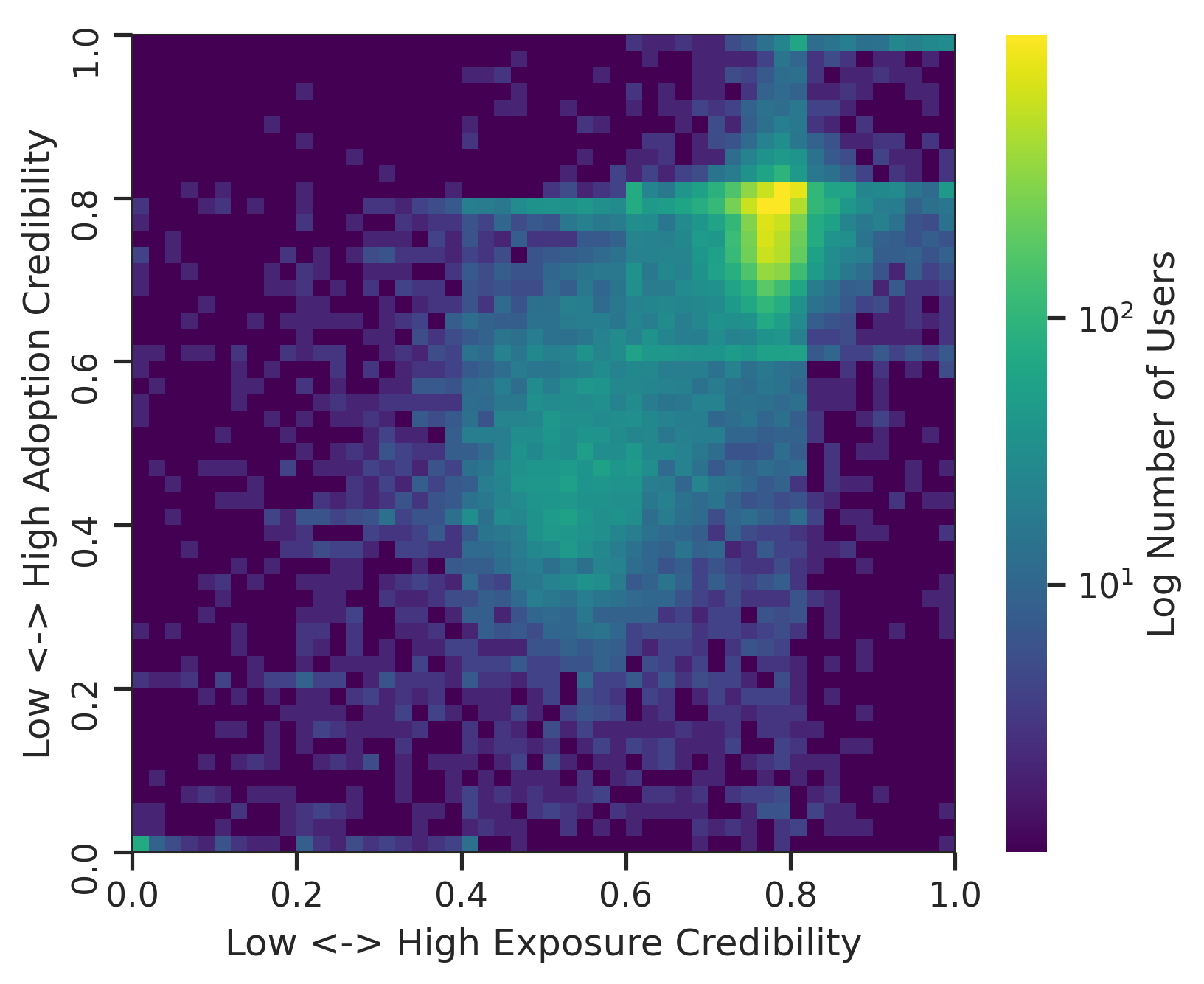}
\caption{Correlation between exposure and adoption credibility scores for the \textsc{Covid} dataset. Pearson's correlation $r$ = 0.66, $p$ < .001.}
\label{heatmap-covid}
\end{figure}

\begin{figure}[H]
\centering
\includegraphics[width=\columnwidth]
{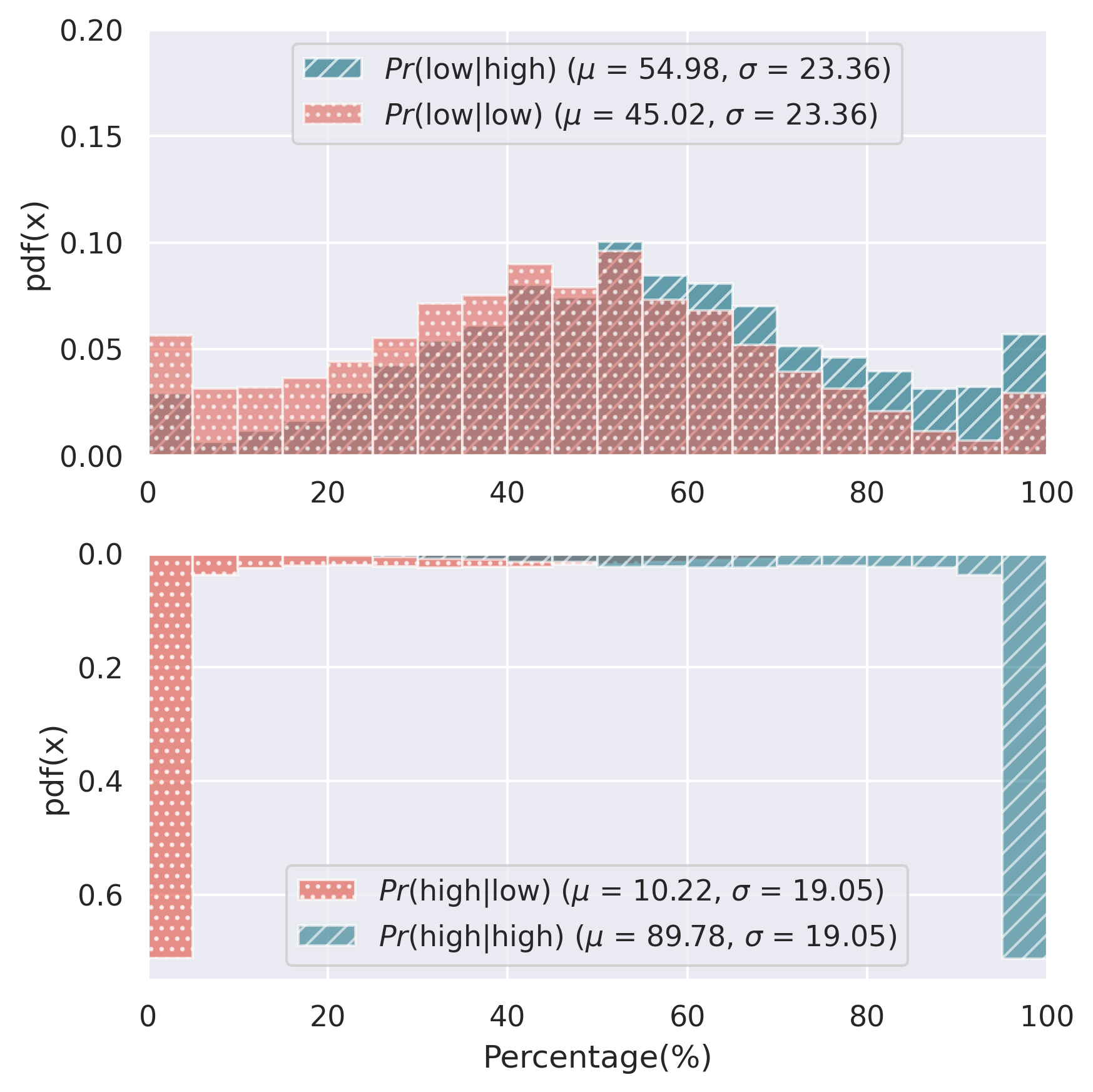}
\caption{Distribution of the percentages of low- and high-credibility exposures in the 7 days prior to low- and high-credibility adoptions in the (\textsc{Covid} dataset).}
\label{bayesian-plot-covid}
\end{figure}

Figure \ref{heatmap-covid} shows the correlation between exposure and adoption credibility in the \textsc{Covid} dataset. 
Figure \ref{bayesian-plot-covid} demonstrates the distribution of the percentages of low- and high-credibility exposures in the 7 days prior to low- and high-credibility adoptions in the \textsc{Covid} dataset.
As part of a robustness check for RQ3 (\S\ref{rq3}), we explore variations using different prior exposure windows (7 days, 3 days, and 1 day), as detailed in Table \ref{bayesian-framework-all}. Using pairwise one-sided Mann-Whitney U tests, we verify that the following distributions are significantly different ($p<.001$) across both datasets: $Pr(low|low)>Pr(high|low)$, $Pr(high|high)>Pr(low|high)$, and $Pr(high|high)>Pr(low|low)$.

\begin{table}[H]
\centering
\small
\caption{Probability of low- and high-credibility exposures prior to low- and high-credibility adoptions.}
\label{bayesian-framework-all}
\begin{tabular}{@{}lcccccc@{}}
\toprule
 & \multicolumn{3}{c}{\textsc{Election} Dataset} & \multicolumn{3}{c}{\textsc{Covid} Dataset} \\
\cmidrule(lr){2-4} \cmidrule(lr){5-7}
Exposure Window & 7 days & 3 days & 1 day &  7 days & 3 days & 1 day \\
\midrule 
$\mu_{Pr(low|high)}$ & 0.523 & 0.517 & 0.572 & 0.550 & 0.542 & 0.528 \\
$\mu_{Pr(low|low)}$ & 0.477 & 0.483 & 0.428 & 0.450 & 0.458 & 0.472 \\
$\mu_{Pr(high|low)}$ & 0.233 & 0.255 & 0.247 & 0.102 & 0.112 & 0.101 \\
$\mu_{Pr(high|high)}$ & 0.767 & 0.745 & 0.753 & 0.898 & 0.888 & 0.899 \\
\bottomrule
\end{tabular}
\end{table}

\end{document}